\documentstyle[prb,amssymb,aps,twocolumn,epsfig,psfig]{revtex}

    \def \a{\alpha}
    
\def \s{\sigma}    \def \k{\kappa} 
\def \e{\epsilon}   
\def \th{\theta}   \def \d{\delta}
\def \k{\kappa}

\def \h{\hbar}  

\def \del{\partial}    % for writing partial derivatives

\def \ra{\rightarrow}

\def\be{\begin{equation}}
\def\ee{\end{equation}}
\def\bea{\begin{eqnarray}}
\def\eea{\end{eqnarray}}
\def\bea*{\begin{eqnarray*}}  
\def\eea*{\end{eqnarray*}}
\def\v{\mathbf}

\begin{document}
\twocolumn[\hsize\textwidth\columnwidth\hsize\csname  
@twocolumnfalse\endcsname

\title{Multi-patch model for transport properties of cuprate superconductors}
\author{A. Perali, M. Sindel, G. Kotliar\cite{email}}
\address{Department of Physics and Astronomy, Rutgers University,\\
136 Frelinghuysen Rd., Piscataway, New Jersey, 08854-8019}
\date{\today}
\maketitle
 
\begin{abstract}
A number of normal state transport properties of cuprate superconductors are analyzed in detail 
using the Boltzmann equation. The momentum dependence of the electronic structure and 
the strong momentum anisotropy of the electronic scattering are included 
in a phenomenological way via a multi-patch model. The Brillouin zone and the Fermi
surface are divided in regions where scattering between the electrons is strong and
the Fermi velocity is low (hot patches) and in regions where the 
scattering is weak and the Fermi velocity is large (cold patches).
We present several motivations for this phenomenology starting from various microscopic
approaches. 
A solution of the Boltzmann equation in the case of N patches is obtained and an expression
for the distribution function away from equilibrium is given.
Within this framework, and limiting our analysis to the two patches case,
the temperature dependence of
resistivity, thermoelectric power, Hall angle, magnetoresistance and thermal Hall conductivity are 
studied in a systematic way analyzing the role of the patch geometry and the temperature dependence
of the scattering rates. In the case of $Bi$-based cuprates, using ARPES data for the electronic structure, 
and assuming an inter-patch scattering between hot and cold states with a linear temperature dependence, a reasonable
agreement with the available experiments is obtained. 
\end{abstract}
]

\section{Introduction}

The normal state transport properties of cuprate superconductors have attracted enormous attention.
In the underdoped regime a pseudogap appears
in the excitation spectrum of the metallic state above the superconducting critical temperature $T_c$
and below a doping dependent crossover temperature $T^{*}$. At optimum doping the $T^{*}$ almost coincides 
with $T_c$ and the pseudogap region of the phase diagram, if present, is very narrow. On the other hand the metallic state
of cuprates at optimum doping, away from the pseudogap region, still 
strongly deviates from what is observed in simple metals.
At optimum doping the in-plane DC-resistivity is linear in temperature from $T_c$ to very high temperatures \cite{Forro1990,Takagi}, 
the thermoelectric power is linear \cite{McIntosh1996}, the cotangent of the Hall angle displays a $T^{\gamma}$-dependence, 
($\gamma \simeq 2$ in $La$-based and $Y$-based cuprates \cite{Chien,Marta}, with deviations at low temperature of the order of $2T_c$, 
$1.60\le \gamma \le 2$ in $Bi$-based cuprates \cite{Konstantinovic2000,Ando1999}, where increasing the
doping decreases the exponent $\gamma$), the magnetoresistance has approximately a $T^{-\alpha}$ dependence, 
with $\alpha\simeq 4$ \cite{Ando1999} and the thermal Hall conductivity has approximately a $T^{-\beta}$ dependence, 
with $\beta\simeq 1.2$ \cite{Zhang2000}.  
Increasing the doping, in the overdoped regime, the conventional Fermi liquid character of the metallic state is almost recovered.

Theoretical approaches to the problem of transport properties of cuprates can be classified in:
({\em i}) approaches where in-plane transport properties are analyzed in terms of two scattering times, one 
associated to the response to an electric field (determining DC-conductivity) and the other one 
associated to the response to a magnetic field (determining  Hall conductivity and magnetoresistance) \cite{Anderson};
({\em ii}) approaches based on the Boltzmann theory \cite{Abrikosov}, where the scattering time is momentum dependent 
and the Fermi surface can be divided in {\em hot} regions (around the $M$ points of the Brillouin zone (BZ)) corresponding 
to strong scattering between quasiparticles (short scattering time) and low Fermi velocity and {\em cold} regions 
(around the nodal points located
along the diagonals $\Gamma Y$ and $\Gamma X$ of the BZ), corresponding to weak 
scattering (large scattering time) and large Fermi velocity.
In the various models \cite{Millis1998,Zheleznyak1998,Hlubina1995} proposed within the class ({\em ii}), the temperature 
dependence of the scattering time in the hot regions strongly deviates from the 
$T^2$ behavior of simple metals, while recovering this conventional behavior in the cold regions.
Hot/cold regions (or spots) models are able to capture some anomalous properties of cuprates, but a general consensus and a 
systematic analysis of the full set of electric and thermal transport properties is lacking in the literature.

In this paper we study in a {\em systematic way} the normal state transport properties of 
cuprate superconductors within the second approach. We introduce a new parameterization of the scattering matrix in the Boltzmann 
equation (BE) for the quasiparticle distribution function via a {\em multi-patch model}, motivated by 
the strong momentum dependence of the electronic properties observed in the cuprates in
angle resolved photoemission spectroscopy (ARPES). 
The BE is solved in the case of N patches and an expression for the perturbed quasiparticle distribution
function is obtained. Our analysis is then limited for simplicity to the case of two patches, leaving the
multi-patch problem for future investigations.
Within the two-patch model the BZ and the Fermi surface (FS)
are divided into two regions, the {\em{hot}} regions corresponding to hot patches on the FS  
and the {\em{cold}} regions corresponding to cold patches on the FS \cite{notatwogap}. Assuming 
a $T^2$ temperature dependence for the scattering amplitude in the cold region, a $T$ dependence for the inter-patch 
(hot-cold) scattering 
and a constant temperature dependence in the hot region, a reasonable description of the available experimental data for
$Bi$-based cuprates is obtained. 

ARPES {\em experiments}, mainly performed in $Bi2212$ compounds, give a qualitative justification to the division of the 
BZ and of the FS in cold and hot regions. ARPES clearly shows that
the line-shape of the spectral function is strongly momentum dependent \cite{Ding,Mesot}. Around the $M$ points of the 
Brillouin zone $(\pm\pi,0)$; $(0,\pm\pi)$ the spectral function is very broad (the line-width is of the order of $0.2-0.3eV$ 
at $T=100 K$ \cite{Mesot})
and a quasiparticle peak cannot be easily distinguished; the states around the $M$ points are therefore almost 
incoherent (as localized states) and a very strong scattering mechanism is at the origin of the broad line-shape. 
Moreover the electronic band dispersion has saddle points located at the $M$ points and this originates the van Hove 
singularity in the density of states. The band dispersion along the $MY$ direction is very narrow
($\Delta \e \approx 50meV$ for optimally doped $Bi2212$) and the Fermi velocity is low \cite{Campuzano}.
These states correspond therefore to {\em hot states}. The incoherent behavior and the associated line-shape 
is also temperature independent in the wide range of temperature between $T_c$ and $300K$. On the other hand, 
around the nodal points of the BZ located along the $\Gamma Y(X)$ diagonals, the spectral function has a well
pronounced quasiparticle peak (the line-width is of the order of $0.05-0.1eV$ at $T=100 K$ \cite{Mesot}) 
and the wave-vector dispersion of this peak together with the temperature dependence 
of the peak width can be followed. Valla {\em et al.} found a linear temperature and frequency dependence of the peak width for
states around the nodal points \cite{Valla}. Moreover the band dispersion along 
$\Gamma Y (X)$ is wide ($\Delta \e \approx 400meV$ for optimally doped $Bi2212$) 
and the Fermi velocity is high. The ratio of the Fermi velocities in the two regions
is $v_F(\Gamma Y)/v_F(M)\simeq 3$ \cite{Marshall}.
The quasiparticle states around the nodal points are therefore 
coherent (delocalized states) and a scattering mechanism with weaker intensity (even with unconventional nature)
is at the origin of the line-shape behavior. These states correspond to {\em cold states}. 

The physics of cuprates is very rich.
Starting from the low doping insulating phase, the holes added to the $CuO_2$ planes of cuprates through 
out-of-plane doping will become metallic and segments of a FS appear. The holes move in 
an antiferromagnetic background and two holes with opposite spins on the same lattice site experience 
a strong Hubbard repulsion. The conducting holes are therefore strongly correlated and a possible description
of the electronic properties of the metallic state of cuprates has been formulated in terms 
of the two dimensional Hubbard model, eventually with the inclusion of other terms in the Hamiltonian
(extended Hubbard model) to take into account short range attraction due to phonons and/or long range Coulomb interaction
which can be relevant at high doping. First insights in the phase diagram (temperature vs. doping) of the (extended)
Hubbard model indicate the presence of several 
electronic instabilities arising from the competition between the different degrees of freedom
present in the Hamiltonian, such as the kinetic term (delocalizing term) and the short range terms (localizing terms).
The most relevant phases, toward which the system is unstable, are the antiferromagnetic insulating phase, 
phase separation in macroscopic regions with low and high density of holes, spin and charge ordering in the 
form of stripes, and finally superconductivity. 
The metallic state of cuprates, in particular in the underdoped region of the phase diagram, can be close to one of 
these electronic instabilities. Therefore, the properties of the metallic state of cuprates can be strongly connected 
to the presence of several competing interactions which arise nearby the electronic instabilities mentioned above. 

There are two possible {\em microscopic origins} for the momentum differentiation of the BZ,
one connected to electronic scattering mediated by spin, charge or pair fluctuations and another associated
with proximity to a Mott transition.

The first possibility for the momentum differentiation of the BZ is based on
its connection to electronic scattering due to spin fluctuations \cite{Carrington}. Superconducting fluctuations
\cite{Millis1998} and charge instabilities \cite{clc95} have also been involved.
These fluctuations can mediate the electron-electron interaction in both
the particle-hole and particle-particle channels and hence they can determine strong deviations in 
the properties of the metallic state, such as the line-shape of the ARPES spectra and the shape of the FS respect
to a conventional Fermi liquid picture.
The various fluctuations have a specific momentum and frequency dependence and the propagators are peaked at different
critical momenta $\v{q}_c$, depending from the underlying instability: for phase separation
$\v{q}_c =0$ \cite{clc95}, for antiferromagnetic insulator $\v{q}_c=\v{Q} \equiv (\pi,\pi)$ \cite{Pines}, for charge ordering
$\v{q}_c=\v{q}_{stripe}$ \cite{Becca,clc96}, where $q_{stripe}=2\pi/\lambda_{stripe}$ ($\lambda_{stripe}$ is the periodicity
of the charge modulation). Therefore, the critical fluctuations couple electrons with momenta 
only inside particular segments of FS, and the specific geometry of the strongly coupled
(and hence of the weakly coupled) segments is determined by the interplay between the shape of the FS and $\v{q}_c$.
This leads to the phenomenology of the two- or multi-patch model for the electronic scattering in the 
metallic state of cuprates. 

The second scenario for momentum differentiation invokes the proximity to a Mott transition.
It builds on our recent understanding of this phenomena within dynamical mean field theory (DMFT) \cite{Kotliarrev} 
and its extension: the two impurity method \cite{Kotliarrev,Schiller}, the Bethe-Peierls cluster \cite{Kotliarrev},
the dynamical cluster approximation (DCA) \cite{Hettler,Jarrell,Huscroft} and 
the cellular dynamical mean field theory (C-DMFT) \cite{Lichtenstein,kotclust,biroliclust}.

DMFT allows a microscopic description of the strongly correlated state near the Mott transition.
There is a temperature scale reminiscent of the Kondo temperature $T_K$, such that for  $T\ll T_K$ the quasiparticles are
Fermi liquid like, while for $T>T_K$ the single particle excitations become incoherent and the transport properties are
non Fermi liquid like. In single site DMFT studies, the two regimes were obtained by varying the temperature and the
strength of the local Hubbard repulsion $U$, and by construction
they occur uniformly in the BZ. Relaxing the constraint of momentum independent selfenergy by using the
C-DMFT, one envisions that different patches of BZ have different Kondo temperatures, leading naturally to the
multi-patch model presented here. According to this view, therefore,
the microscopic origin of the multi-patch division of the BZ and of the FS
of cuprate superconductors can be associated to the presence of a nearby transition from a (non conventional) 
metallic state to a Mott antiferromagnetic insulator. 
A detailed microscopic derivation of the many-patch model is in progress, using the C-DMFT.

These views on momentum space differentiation may be complementary, and there are already strong hints
from numerical calculations by Onoda and Imada \cite{Onoda}, that they occur in the Hubbard model.

The plan of the paper is the following. In Section \ref{s2} we recall the BE in the linearized form and we introduce 
the multi-patch model for the the scattering operator. 
The scattering operator of the BE is projected on the patches and temperature dependences of its coefficients 
are assigned. A set of smooth functions is introduced to permit a continuous transition between hot and cold regions. 
A solution of the BE in the case of N patches is given in terms of the perturbed quasiparticle distribution
function. In Section \ref{s3} the results for the normal state transport properties 
obtained by the two-patch model are presented in a {\em systematic way}. 
The temperature dependences of resistivity, thermoelectric power, cotangent of the Hall angle, 
magnetoresistance and thermal Hall conductivity are reported. 
The various transport properties are studied for different set of parameters and the relevant hot/cold
patch is associated to every quantity. Our model is  
applied to the normal state transport properties of $Bi$-based cuprates ($Bi 2212$ and $Bi 2201$) and, taking ARPES data 
as input for the electronic structure, we find a reasonable agreement with the 
available experimental data. Discussions and conclusions are given in Section \ref{s5}.

\section{The Boltzmann equation and the multi-patch model} \label{s2}

ARPES experiments show that optimally and overdoped cuprates have a large FS 
in the normal state and well defined quasiparticles exist in a sizeable (cold) region of the BZ around the $\Gamma Y(X)$
directions. The existence of a FS and quasiparticles
makes the treatment within the framework of the BE possible. We start our analysis introducing 
the linearized BE. As we are interested in electrical, heat and Hall transport properties, we consider three terms in the BE,
the terms including the electric field $\v{E}$ and the temperature gradient $\nabla_{\v{r}} T/T$ (driving terms) and 
the term including the magnetic field $\v{B}$ (bending term). 
We consider here uniform electric and magnetic field.
In this case the linearized
BE has the following form:
\begin{eqnarray} \label{linBoltz}
\nonumber \frac{\del}{\del t}g_{\v{k}}&+&
\left(-\frac{\del f_{\v{k}}^0}{\del \e_{\v{k}}}\right) \e_{\v{k}}  
{\v{v}}_{\v{k}}\cdot\frac{\nabla_{\v{r}} T}{T} \\
&+&e{\v{E}}{\v{v}}_{\v{k}}\left(\frac{\del f_{\v{k}}^0}{\del \e_{\v{k}}}\right)
+\frac{e}{\hbar c}[{\v{v}}_{\v{k}}\times{\v{B}}] \cdot \frac{\del \tilde{g}_{\v{k}}}{\del \v{k}}=C_{\v{k}}
\end{eqnarray}
where $\tilde{g}_{\v{k}}=g_{\v{k}}+(-\del f_{\v{k}}^0)/(\del \e_{\v{k}})\sum_{\v{k}'}f_{\v{k},\v{k}'}
g_{\v{k}'}$ takes the interaction between quasiparticles $f_{\v{k},\v{k}'}$ into account 
(Fermi liquid corrections) \cite{Pinesbook} and 
$g_{\v{k}}$ is the departure from the equilibrium distribution function $f_{\v{k}}^0$. 
${\v{v}_{\v{k}}}$ is the group 
velocity of the quasiparticles.
The scattering operator $C_{\v{k}}$ has the form
$C_{\v{k}}=\sum_{\v{k}'}\left[C_{\v{k},\v{k}'}\tilde{g}_{\v{k}'}-C_{\v{k},\v{k}'}\tilde{g}_{\v{k}}\right]$, 
where $C_{\v{k},\v{k}'}$ is the scattering matrix, describing the scattering of quasiparticles on an 
effective bosonic mode or impurity centers
with the first term describing scattering "in" to the state $\v{k}$ and the second term describing 
scattering "out" of the state $\v{k}$.
The relaxation time $\tau_{\v{k}}$ for the state $\v{k}$
is defined as $1/\tau_{\v{k}}\equiv\sum_{\v{k}'}C_{\v{k},\v{k}'}$.
Within the conventional microscopic approach to the transport properties of
a Fermi liquid, the transport equation for the distribution function can be written
in terms of the four-point vertex part $\Gamma({\v{k},\v{k}',\v{q},\omega})$ \cite{Eliashberg}.
The scattering matrix $C_{\v{k},\v{k}'}$ is then identified with the $\v{q}=0$, $\omega=0$
limit of the irreducible part of the vertex $C_{\v{k},\v{k}'}=$i$(z^2/2)\Gamma^{(1)}({\v{k},\v{k}',\v{q}=0},\omega=0)$,
where $z^2$ is the mass renormalization and $\Gamma^{(1)}$ is the irreducible vertex. 
Below we will use this identification to give a first 
microscopic justification to our choice of $C_{\v{k},\v{k}'}$.
In the steady-state case we can  replace  $\tilde{g}_{\v{k}}\ra g_{\v{k}}$ 
as every term in Eq. (\ref{linBoltz}) is expressed in terms of $\tilde{g}_{\v{k}}$. 
Therefore the knowledge of the form of $f_{\v{k},\v{k}'}$ is not required in the steady-state case.
Frequency-dependent transport processes require both $\tilde{g}_{\v{k}}$ and $g_{\v{k}}$, 
and hence the knowledge of $f_{\v{k},\v{k}'}$ becomes important.
In the steady-state case the number of quasiparticles $g_{\v{k}}$ contributing to transport is given by
\begin{eqnarray} \label{bol}
\left[\frac{e}{\hbar c}({\v{v}}_{\v{k}}\times\v{B})\cdot\nabla_{\v{k}}\right.&+&
\left.\frac{1}{\tau_{\v{k}}}\right]
g_{\v{k}}
-\sum_{\v{k}'}C_{\v{k},\v{k}'}g_{\v{k}'}= \\ \nonumber &=& \left[e{\v{v}}_{\v{k}}\cdot {\v{E}}-
 \e_{\v{k}}{\v{v}}_{\v{k}}\cdot\frac{\nabla_{\v{r}}T}{T}\right]
\left(-\frac{\del f_{\v{k}}^0}{\del \e_{\v{k}}}\right). 
\end{eqnarray}
Defining the operator $\hat{A}_{\v{k},\v{k}'}$ as 
\be \label{Ak}
\hat{A}_{\v{k},\v{k}'}\equiv 
\left[\frac{1}{\tau_{\v{k}}}+\frac{e}{\hbar c}(\v{v}_{\v{k}}
\times\v{B})\cdot\nabla_{\v{k}}\right]\d_{\v{k},\v{k}'}-C_{\v{k},\v{k}'}
\ee
allows us to write the l.h.s. of  Eq. (\ref{bol}) in the form $\sum_{\v{k}'}\hat{A}_{\v{k},\v{k}'}
g_{\v{k}'}$ and hence the inverse of $\hat{A}_{\v{k},\v{k}'}$ is required to
solve Eq. (\ref{bol}).  
Considering weak magnetic fields, the bending term containing the magnetic field in the BE can be treated  
 as a small perturbation
of the transport process. We split $\hat{A}_{\v{k},\v{k}'}$ in two parts, 
$\hat{A}_{\v{k},\v{k}'}=\hat{K}_{\v{k},\v{k}'}+\hat{M}_{\v{k},\v{k}'}^B$, with 
a magnetic field independent part $\hat{K}_{\v{k},\v{k}'} = (1/\tau_{\v{k}})\d_{\v{k},\v{k}'}-C_{\v{k},\v{k}'}$ 
and 
a part that contains the magnetic field  
$\hat{M}_{\v{k},\v{k}'}^B=
\left[(e/\hbar c)(\v{v}_{\v{k}}\times\v{B})\cdot\nabla_{\v{k}}\right]\d_{\v{k},\v{k}'}$.
A perturbative expansion of $\hat{A}_{\v{k},\v{k}'}$ in powers of $\v{B}$ allows us to write 
the inverse of this operator as  
\begin{eqnarray}\label{Ainverse}
\hat{A}_{\v{k},\v{k}'}^{-1}&=&
\hat{K}_{\v{k},\v{k}'}^{-1}-\hat{K}_{\v{k},\v{k}^i}^{-1}\hat{M}_{\v{k}^i,\v{k}^j}^{B}
\hat{K}_{\v{k}^j,\v{k}'}^{-1}\\ \nonumber &+&\hat{K}_{\v{k},\v{k}^i}^{-1}
\hat{M}_{\v{k}^i,\v{k}^j}^{B}\hat{K}_{\v{k}^j,\v{k}^k}^{-1}\hat{M}_{\v{k}^k,\v{k}^l}^{B}
\hat{K}_{\v{k}^l,\v{k}'}^{-1}
+{\mathcal{O}}(B^{3})
\end{eqnarray} 
with a summation over repeated indexes.
Depending on the quantity of interest, we get contribution  from the different terms in this expansion.
The first term gives the leading order contribution to the DC and thermal conductivity, the second term 
to the (thermal) Hall-conductivity and the third term to the magnetoresistance.
It follows from Eq. (\ref{bol}) that the number of particles $g_{\v{k}}$ contributing to transport is given by
\be  \label{freeparticles}
g_{\v{k}}=\sum_{\v{k}'}\hat{A}^{-1}_{\v{k},\v{k}'}\left[e{\v{v}_{\v{k}'}}\cdot {\v{E}}-
 \e_{\v{k}'}{\v{v}_{\v{k}'}}\cdot\frac{\nabla_{\v{r}}T}{T}\right] 
\left(-\frac{\del f_{\v{k}'}^0}{\del \e_{\v{k}'}}\right).
\ee
This equation is the starting point of our analysis. Electrical and thermal conductivities 
are derived using $g_{\v{k}}$ obtained from Eq. (\ref{freeparticles}).

The transport properties are separable in electric and thermal properties. 
All possible currents are given by
\be \label{posscurrents}
\left(\begin{array}{c}\v{j}_e\\ \v{j}_Q \end{array}\right)=\left(\begin{array}{cl}\bar{\sigma} & \bar{S} \\ \bar{S} & 
\bar{\kappa}\end{array}\right)
\left(\begin{array}{c}\v{E} \\ -\frac{\nabla T}{T}\end{array}\right)
\ee
where ${\v{j}}_e$ is the electric current, $\v{j}_{Q}$ is the thermal current, 
$\bar{\sigma}$ is the electrical conductivity tensor, 
$\bar{\kappa}$ is the thermal conductivity tensor and $\bar{S}$ is the thermopower tensor.  
Note that Eq. (\ref{posscurrents}) contains a symmetric matrix using $\nabla T/T$ as driving term for 
thermal gradients. The currents defined in Eq. (\ref{posscurrents}) can also be expressed 
in terms of quasiparticles. The electric and the thermal 
currents are given by ${\v{j}}_e=e\sum_{\v{k}}{\v{v}}_{\v{k}}g_{\v{k}}$ and 
${\v{j}}_Q=\sum_{\v{k}}{\v{v}}_{\v{k}}\e_{\v{k}}g_{\v{k}}$. 
(When frequency dependence is considered, $g_{\v{k}}$ has to be replaced with
$\tilde g_{\v{k}}$ in the expression of the currents given above.)
The tensors defined in 
Eq. (\ref{posscurrents}) are given by:
\be \label{electricconduc}
\sigma^{\mu \nu}=2e^2\sum_{\v{k},\v{k}'}v^{\mu}_{\v{k}} \hat{A}^{-1}_{\v{k},\v{k}'} v^{\nu}_{\v{k}'}
\left(-\frac{\del f_{\v{k}'}^0}
{\del \e_{\v{k}'}}\right)
\ee
\be \label{thermconduc}
\k^{\mu\nu}=2\sum_{\v{k},\v{k}'}v^{\mu}_{\v{k}} \e_{\v{k}}
\hat{A}^{-1}_{\v{k},\v{k}'} \e_{\v{k}'}v^{\nu}_{\v{k}'}\left(- \frac{\del f_{\v{k}'}^0}{\del \e_{\v{k}'}}
\right)
\ee
\be \label{thermopower}
S^{\mu\nu}=-2e\sum_{\v{k},\v{k}'}v^{\mu}_{\v{k}} \e_{\v{k}}
\hat{A}^{-1}_{\v{k},\v{k}'} v^{\nu}_{\v{k}'}\left(- \frac{\del f_{\v{k}'}^0}{\del \e_{\v{k}'}}
\right)
\ee
with the inverse of the operator $\hat{A}_{\v{k},\v{k}'}$ given in Eq. (\ref{Ainverse}). The factor $2$
in the expressions above takes the spin degeneracy into account. 

The last unknown quantity in Eq. (\ref{bol}) is the scattering matrix $C_{\v{k},\v{k}'}$.
The scattering matrix is connected via a frequency integration to the spectral function of
the effective bosonic mode exchanged in the electronic scattering. As discussed above, ARPES
suggests that the bosonic mode is strongly momentum dependent and divides the BZ in hot and
cold regions. In order to include in the scattering matrix a non trivial momentum dependence
and the cold/hot division of the BZ, while maintaining a simple solution of the BE,
a possibility is to expand $C_{\v{k},\v{k}'}$ with respect to a basis of functions which are 
able to select the various regions of the BZ, according to the momentum dependence of the
effective bosonic mode. Therefore, the scattering matrix $C_{\v{k},\v{k}'}$ can be written as
\begin{equation}
\label{expckk}
C_{\v{k},\v{k}'}=\sum_{i,j=1}^{N}a_{ij}\Phi_i({\v{k}}) \Phi_j({\v{k}'}),
\end{equation}
where $\Phi_i(\v{k})$ is a function which is equal to one inside the $i$-th patch of the BZ
and zero outside, and it interpolates continuously between these two values;
$a_{ij}$ is the amplitude of the scattering between the $i$-th and the $j$-th region of the BZ,
it is in general temperature dependent and $((a_{ij}))$ is a symmetric matrix with real elements; N is the total
number of regions of the BZ required to include properly the main effects of the anisotropy
of the scattering and its interplay with the shape of the FS.

One possible {\em microscopic motivation} for the form of the scattering matrix given in Eq. (\ref{expckk}) is
based on the connection between $C_{\v{k},\v{k}'}$ and the four-point vertex $\Gamma$ discussed 
above Eq. (\ref{bol})
and on the C-DMFT. The four-point vertex can in principle be evaluated by C-DMFT
and the general structure of the irreducible vertex will be
$\Gamma^{(1)}({\v{k}},{\v{k}'},{\v{q}},\omega)=\sum_{\alpha,\beta,\gamma,\delta}\Gamma^c_{\alpha,\beta,\gamma,\delta}(\omega)
\phi_{\alpha}^{\dag}({\v{k}})\phi_{\beta}({\v{k}'})\phi_{\gamma}^{\dag}({\v{k}+\v{q}})\phi_{\delta}({\v{k}'+\v{q}})$.
Once the patch function is defined as $\Phi_{i=(\alpha,\beta)}({\v{k}})
=\phi_{\alpha}^{\dag}({\v{k}})\phi_{\beta}({\v{k}})$,
Eq. (\ref{expckk}) for $C_{\v{k},\v{k}'}$ can be recovered.
The coefficient $a_{ij}$ entering Eq. (\ref{expckk}) are then defined in terms of the
four-point vertex of the cluster as $a_{i=(\alpha,\beta),j=(\gamma,\delta)}=$i$(z^2/2)\Gamma^c_{\alpha,\beta,\gamma,\delta}(\omega=0)$.

The form of the scattering matrix given in Eq. (\ref{expckk}) permits an {\em analytical solution} 
of the linearized BE for $g_{\v{k}}$ defined above. Considering the electric field and the thermal gradient
as external perturbations, Eq. (\ref{bol}) gives 
\begin{eqnarray}
\label{mpgk}
g_{\v{k}}=\tau_{\v{k}}\left[e{\v{v_k}}{\v{E}}-\e_{{\v{k}}}{\v{v_k}}\frac{\nabla_{{\v{r}}} T}{T}\right]
\left(- \frac{\del f_{{\v{k}}}^0}{\del \e_{{\v{k}}}}\right)
+\tau_{{\v{k}}}\sum_{i,j=1}^N a_{ij} \Phi_i({\v{k}})\cdot \\
\nonumber 
\cdot \sum_{\v{k'}}\left[e{\v{v_k}}{\v{E}}-\e_{{\v{k}}}{\v{v_k}}\frac{\nabla_{{\v{r}}} T}{T}\right]
\left(- \frac{\del f_{\v{k'}}^0}{\del \e_{\v{k'}}}\right) 
\sum_{l=1}^N T^{-1}_{jl}\Phi_l({\v{k'}}),
\end{eqnarray}
which reduces the solution of the BE to a matrix inversion problem for the N$^2$ elements of
a matrix $T_{jl}$ given by

\begin{equation}
\label{tij}
T_{jl}=\delta_{jl}-\sum_{m=1}^N a_{mj}\sum_{\v{k}}\Phi_m({\v{k}})\tau_{\v{k}}\Phi_l({\v{k}}).
\end{equation}

From Eq. (\ref{mpgk}) and Eq. (\ref{tij}) we can obtain the inverse of the
matrix $\hat{K}_{\v{k},\v{k}'}$, defined above Eq. (\ref{Ainverse}), as follow

\begin{equation}
\label{kkk}
\hat{K}^{-1}_{\v{k},\v{k}'}=\tau_{\v{k}}\delta_{\v{k},\v{k}'}+\tau_{\v{k}}\tau_{\v{k'}}
\sum_{i,j=1}^N a_{ij} \Phi_i({\v{k}})
\sum_{l=1}^N T^{-1}_{jl}\Phi_l({\v{k'}}) .
\end{equation}

Once we know the explicit expression for the operator $\hat{K}^{-1}_{\v{k},\v{k}'}$, we are able to evaluate
the expansion of the operator $\hat{A}^{-1}_{\v{k},\v{k}'}$ for weak magnetic fields given in Eq. (\ref{Ainverse}).
Inserting the operator $\hat{A}^{-1}_{\v{k},\v{k}'}$ in Eq. (\ref{freeparticles}), we can evaluate the solution
of the linearized BE, given by $g_{\v{k}}$, in the presence of an electric field, a thermal gradient
and a weak magnetic field. $g_{\v{k}}$ can be then used to evaluate all the currents given above
Eq. (\ref{electricconduc}). All the response functions given in Eqs. (\ref{electricconduc}, \ref{thermconduc}, \ref{thermopower})
are obtained inserting directly the expression for $\hat{A}^{-1}_{\v{k},\v{k}'}$.

In the following we consider a minimal realization of this multi-patch approximation 
limiting our analysis to N=2 patches. The {\em two-patch model} permits the distinction between cold and hot
regions on the BZ and the FS and it is suitable to describe the main properties of the scattering
between electrons and an effective mode peaked at large momenta ($e.g.$ an antiferromagnetic spin
fluctuation peaked at $\v{Q}\equiv (\pi;\pi)$).
On the other hand, to include the effect of the (small momenta) forward scattering, important for
transport mainly in the cold region, a larger number of patches is required (at least N=5), 
and we deserve this case for further investigation.

\begin{figure} 
\centering \epsfig{file=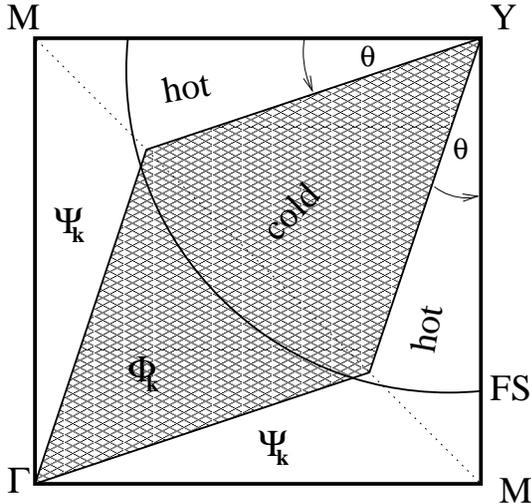, width=7cm}
\caption{The first quadrant of the BZ is splitted into  
cold and  hot regions.
Cold regions are described by a function $\Phi_{\v{k}}$ and hot regions by a function $\Psi_{\v{k}}$.
The angle $\th$ parameterizes the size of the hot regions.}
\label{geotwopatch}
\end{figure}

\begin{figure} 
\centering \psfig{file=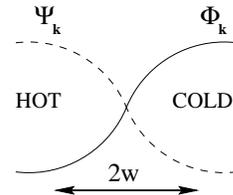, width=3cm}
\caption{Smooth 
change between the two patches. The parameter $w$ describes the width of the 
transition region between the two patches and it allows to study the effects 
of this region on transport properties.}
\label{smooth}
\end{figure}

In the case N=2, the two-patch division of the BZ and of the Fermi surface 
is realized by introducing two functions $\Phi_{\v{k}}$ and $\Psi_{\v{k}}$:
$\Phi_{\v{k}}$ describes
the cold region and $\Psi_{\v{k}}$ the hot region as indicated in Fig. \ref{geotwopatch}. 
The cold region is described by four boundaries as shown in Fig. \ref{geotwopatch}. 
The angle $\th$, defined with respect to the $Y=(\pi,\pi)$ point of the BZ, 
parameterizes the size of the hot regions (and hence of the cold regions).
The reason to split the whole BZ (and not only the FS) into a cold and a hot part is 
that the term $\del f^0/\del \e$ in Eq. (\ref{bol}) doesn't restrict the sum over $\v{k}$ 
on the FS any more when we increase size-ably the temperature, studying $T$-dependent properties.  
As the leading contribution to the magnetoresistance is given by the third term in Eq. (\ref{Ainverse}), which contains 
two derivatives with respect to $k_x, k_y$  (given by $\hat{M}^B$ repeated two times),
the magnetoresistance diverges in the case of a discontinuous change between the two patches. 
Therefore we introduce functions $\Phi_{\v{k}}$ and $\Psi_{\v{k}}$ varying smoothly between 
the two regions. We use hyperbolic tangents to describe the smooth change between the two regions and a 
parameter $w$ is introduced to describe the width of the transition region. 
In the limit $w\ra 0$  
a step function is recovered, $\th(x)=\lim_{w\rightarrow 0}(1+\tanh(x/w))/2$. 
Finally, we define two slopes, $m_1$ and $m_2$, and two offsets,
$t_1$ and $t_2$, given by
$m_1=\cot(\th)>0$, $m_2=\tan(\th)>0, m_1>m_2$, $t_1=\pi(1-\tan(\th))>0$ and $t_2=\pi(1-\cot(\th))<0$. 
In the case 
of smooth functions, the cold region in the first quadrant of the BZ (see Fig. \ref{geotwopatch}) is described by:
\be \label{smoothcold}
\Phi_{\v{k}}=\prod_{i=1}^{4}\Phi_i({\v{k}})
\ee
with
\begin{eqnarray*}
\Phi_1({\v{k}})&=&\frac{1}{2}\left[1- \tanh\left(\frac{k_y-m_1k_x}{w}\right)\right];\\
\Phi_2({\v{k}})&=&
\frac{1}{2}\left[\tanh\left(\frac{k_y-m_2k_x}{w}\right)+1\right];\\
\Phi_3({\v{k}})&=&\frac{1}{2}\left[1- \tanh\left(\frac{k_y-m_2k_x-t_1}{w}\right)\right];\\
\Phi_4({\v{k}})&=&\frac{1}{2}\left[\tanh\left(\frac{k_y-m_1k_x-t_2}{w}\right)+1\right]. 
\end{eqnarray*}
The smooth change between the two patches is shown in Fig. \ref{smooth}.
The hot region is described by $\Psi_{\v{k}}=1-\Phi_{\v{k}}$. 
Note that
$\Psi_{\v{k}}$ vanishes only asymptotically in the cold region and vice versa.

Following the phenomenological approach described above,
the scattering matrix can be written consisting of different scattering mechanisms in different patches. 
In the case of the two-patch model, we introduce
three parameters that describe the possible scattering, scattering inside the 
cold or inside the hot region and scattering between the two (hot/cold) regions,  
\be \label{collisionstep}
C_{\v{k},\v{k}'}=a\Phi_{\v{k}}\Phi_{\v{k}'}+b\Psi_{\v{k}}\Psi_{\v{k}'} +c[\Phi_{\v{k}}\Psi_{\v{k}'}+
\Psi_{\v{k}}\Phi_{\v{k}'}].
\ee
The scattering matrix $C_{\v{k},\v{k}'}$ given in Eq. (\ref{collisionstep}) is written as a sum of terms which have a separate
dependence from $\v{k}$ and ${\v{k}}'$. This approximation permits an analytic solution
of the BE for $g_{\v{k}}$, while having a non trivial momentum dependence and symmetry properties of the scattering process.
With the symmetric scattering matrix given in Eq. (\ref{collisionstep}), we obtain a momentum dependent 
relaxation time $\tau_{\v{k}}$, $1/\tau_{\v{k}}\equiv\sum_{\v{k}'}C_{\v{k},\v{k}'}$, 
\be\label{defrelax}
\tau_{\v{k}}= \frac{1}{C_{\Phi}\Phi_{\v{k}}+C_{\Psi}\Psi_{\v{k}}}
\ee
with $C_{\Phi}=\a a+(1-\a)c$ and $C_{\Psi}=(1-\a)b+\a c$, where, in the limit $w\rightarrow 0$, $\a$ 
describes the area of the cold region, 
$\a=\sum_{\v{k}}\Phi_{\v{k}}$, and $1-\a$ the area 
of the hot region, $1-\a=\sum_{\v{k}}\Psi_{\v{k}}$. 
All the sums are normalized with respect to the number of $k$-points of the BZ. 
In the limit $w\ra0$ we get two different lifetimes in the cold and in the hot regions:
\be\label{relaxtimes}
\lim_{w\ra0}\tau_{\v{k}}= \left\{ \begin{array}{ll} \tau_c= \frac{1}{\alpha a +(1 - \alpha)c} & \mbox{in cold regions}, \\ 
\tau_h=\frac{1}{\alpha c +(1 - \alpha)b} & \mbox{in hot regions}. 
\end{array}  \right.
\ee 
We consider the following temperature dependences 
of the scattering amplitudes in the scattering matrix:
\be \label{Tdepend}
a(T)=\bar{a}T^2; \; b(T)=\bar{b}; \;  c(T)=\bar{c}T,
\ee
with the temperature independent parameters $\bar{a}$, $\bar{b}$ and $\bar{c}$ having 
proper dimensions given by  $\left[\h\bar{a}/k_B^2\right]=1/eV$,
$\left[\h\bar{b}\right]=eV$ and $\left[\h\bar{c}/k_B\right]=1$. 
Indeed, in the cold region the scattering is weak and a Fermi liquid behavior with a $T^2$
temperature dependence of the scattering matrix $C_{\v{k},\v{k}'}$ (with both $\v{k},\v{k}'$ inside
the cold region) is a reasonable assumption.
On the other hand, in the hot region the scattering is strong and, as suggested by ARPES, 
the states are almost incoherent; we consider $C_{\v{k},\v{k}'}$ (with both $\v{k},\v{k}'$ inside
the hot region) as temperature independent. Finally, we introduce a coupling between the hot and 
cold regions. The inter-patches elements of $C_{\v{k},\v{k}'}$ (with $\v{k}$ inside the cold region
and $\v{k}'$ inside the hot region and vice-versa) are considered to be temperature dependent with a linear in
$T$ dependence, and this is a {\em key assumption} in our model to obtain the linear temperature dependence of the
resistivity, as shown in the next section.
The important consequence of our assumptions for $C_{\v{k},\v{k}'}$, and in particular the introduction of 
the inter-patches scattering with a linear temperature dependence, 
is that at low temperature the scattering amplitude is non-Fermi liquid at any ${\bf k}$-point of the
BZ. Indeed in the cold region we have $1/\tau_c=\alpha \bar{a}T^2+(1-\alpha )\bar{c}T$,
while in the hot region $1/\tau_h=(1-\alpha )\bar{b}+\alpha \bar{c}T$.
The Fermi liquid behavior can be recovered in the cold region when the area of the hot region 
tends to zero (i.e. $\alpha=1$). The non Fermi liquid behavior of the scattering amplitude in the cold region
at low temperature is supported by the ARPES experiments of Valla {\em et al.} \cite{Valla}, as already discussed. 

To solve the BE, it is necessary to find how the operator $\hat{K}^{-1}$ acts on an arbitrary velocity, as can be seen 
in Eq. (\ref{freeparticles}). 
Because of the symmetry properties of the scattering matrix 
$C_{\v{k},\v{k}'}$ and of the quasiparticle velocity $\v{v_k}$, when 
summing over the whole BZ, $\sum_{\v{k}'}C_{\v{k},\v{k}'}{\v{v}_{\v{k}'}}=0$ \cite{notamp}.
Using these properties, we find that $\hat{K}^{-1}$ acts on an arbitrary velocity
just by inserting a scattering time $\tau_{\v{k}}$, that is momentum dependent, in front of the velocity,
$\sum_{\v{k}'}\hat{K}_{\v{k},\v{k}'}^{-1} v_{\v{k}'}^{\nu}=\tau_{\v{k}} v_{\v{k}}^{\nu}$.
This allows us to solve exactly the linearized BE and to obtain 
formulas for the several transport properties we are interested in.
The deviation $g_{\v{k}}$ from the equilibrium distribution function $f_{\v{k}}^0$ is
$g_{\v{k}}=eE\tau_{\v{k}}{\v{v}_{\v{k}}}\cdot{\hat{{\bf n}}}(-\del f_{\v{k}}^0/\del \e_{\v{k}})$, 
where ${\hat{{\bf n}}}$ is the direction of the external electric field ${\v{E}}=E{\hat{{\bf n}}}$. The momentum dependence 
of $g_{\v{k}}$ is therefore given by 
$g_{\v{k}} \sim {\v{v}_{\v{k}}}\cdot{\hat{{\bf n}}}/(C_{\Phi}\Phi_{\v{k}}+C_{\Psi}\Psi_{\v{k}})$
and hence in the hot regions $g_{\v{k}}$ is strongly suppressed while in the cold regions $g_{\v{k}}$ has sizeable 
values. Hlubina and Rice, solving the BE by a variational approach in the case of hot spots generated by
antiferromagnetic fluctuations coupled to fermions, obtained a $g_{\v{k}}$ with similar properties, i.e. a
depopulation of the hot regions \cite{Hlubina1995} (see Section IV for a discussion).
Finally, as regard the single particle electronic properties, we recall that optimally and overdoped cuprates 
display very similar shape and momentum dependence of the Fermi surface and of the electronic band dispersion. 
Therefore we consider $Bi2212$ as typical for the FS and band structure of cuprates. 
For this material several ARPES measurements are available.
We consider a tight-binding model for the band structure of $Bi2212$ with hopping up to the fifth nearest neighbors.
We are using the following fit for the energy of the quasiparticles $\e (\v{k})$ obtained
by Norman {\em et al.} \cite{Norman1995}.
\be \label{Norman}
\e (k_x,k_y)=\sum_{i=1}^{6}c_i \eta_i(k_x,k_y)
\ee
with the values of the coefficients $c_i$ and the functions $\eta_i$ given by $c_1=0.1305eV$, 
$c_2=-0.5951eV$, $c_3=0.1636eV$, $c_4=-0.0519eV$,
$c_5=-0.1117eV$, $c_6=0.0510eV$ and $\eta_1=1$, $\eta_2=0.5[\cos(k_x)+\cos(k_y)]$, 
$\eta_3=\cos(k_x)\cos(k_y)$, $\eta_4=0.5[\cos(2k_x)+\cos(2k_y)]$,  
$\eta_5=0.5[\cos(2k_x)\cos(k_y)+\cos(2k_y)\cos(k_x)]$, $\eta_6=\cos(2k_x)\cos(2k_y)$. 
These parameters are appropriate
for the band structure of $Bi2212$ compounds at optimum doping, thus giving an open FS and a van Hove singularity (VHS)
slightly below the Fermi level. 
The value of $\e_F=-c_1=-0.1305eV$ is fixed to have the proper distance of the Fermi level
from the VHS ($\e_F-\e_{VHS}=35meV$ as determined experimentally), and corresponds to
the optimum doping $\d =0.17$. The bandwidth of the dispersion given in Eq. (\ref{Norman}) is $1.4eV$. 

\section{Transport properties} \label{s3}

\subsection{Resistivity}

\begin{figure} 
\centering \psfig{file=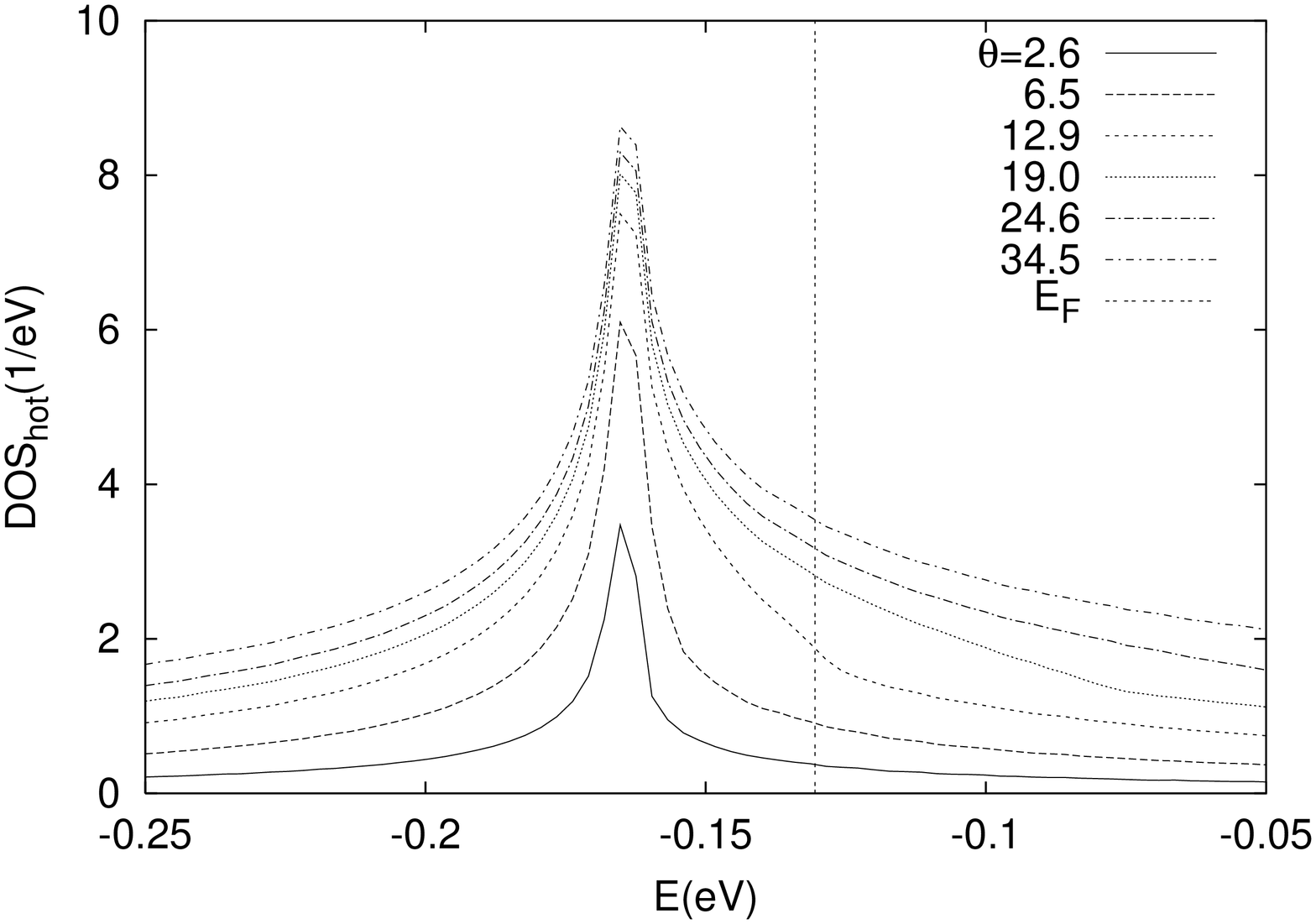, width=6cm, angle=-90}
\centering \psfig{file=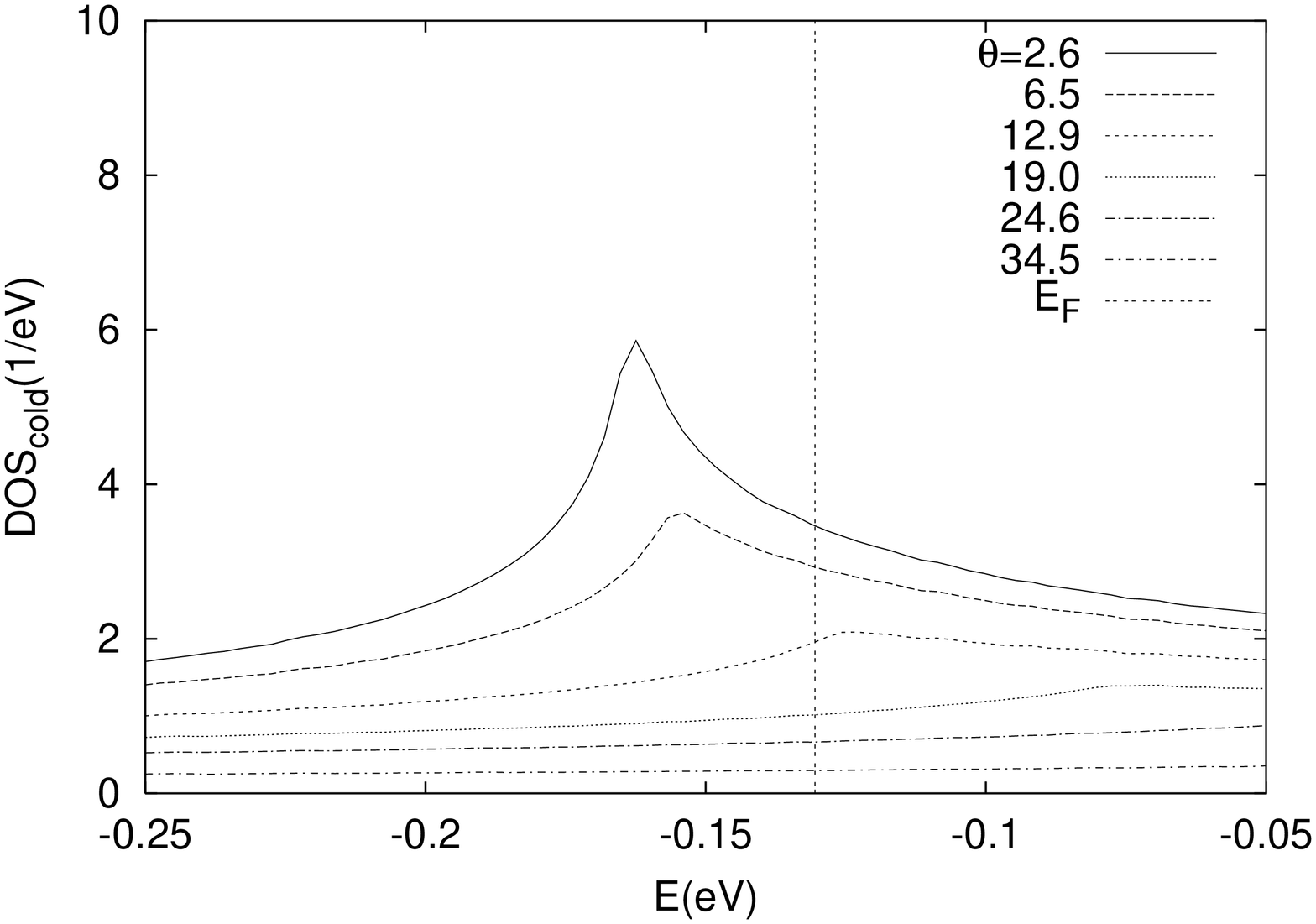, width=6cm,angle=-90}
\caption{The overall shape of the hot density of states $N_h(\e)$ (upper panel), 
doesn't change with $\th$, while the magnitude at the Fermi level changes.
In the case of the cold density of states $N_c(\e_F)$ (lower panel), both the overall shape and the magnitude 
at the Fermi level changes. The position of the Fermi level $E_F$, corresponding to optimum doping, is also
reported. Note the changement of sign in the derivative of $N_c(\e_F)$ for $\th =12.9^{\circ}$.}
\label{figdos}
\end{figure}

{\em Theoretical results.}
The leading order contribution to the DC-conductivity is given by the first term in Eq. (\ref{Ainverse}), thus  
 $\hat{A}^{-1}_{\v{k},\v{k}'}$ is replaced by $\hat{K}^{-1}_{\v{k},\v{k}'}$ and inserted   
into Eq. (\ref{electricconduc}). The operator $\hat{K}^{-1}_{\v{k},\v{k}'}$, applied on a velocity $v_{\v{k}}$
gives $v_{\v{k}}\tau_{\v{k}}$, as already discussed in Section \ref{s2}. Therefore, the DC-conductivity is given by:
\be \label{sigmaxx}
\sigma^{xx}= 2 e^2\sum_{\v{k}}\left(-\frac{\del f_{\v{k}}^0}{\del \e_{\v{k}}}\right)
(v_{\v{k}}^x)^2 \tau_{\v{k}}.
\ee
In the limit $T\ra 0$ we can obtain a finite contribution to $\s^{xx}$ depending on the width $w$ and
on the angle $\th$ (determining the area of the hot region). As
$\tau_{\v{k}}$ given in Eq. (\ref{defrelax}) doesn't diverge in the case $w\neq 0$ (because $\Psi_{\v{k}}\neq 0$ 
even inside the cold region), we get a finite contribution to $\s^{xx}$ and a residual resistivity
at zero temperature. Note that the residual resistivity is determined by the width $w$ and $\bar{b}(1-\a)$, thus the 
$T$-independent part in the coefficient $C_{\Psi}$.  
In the limit $w\ra0$
we separate two regions in the BZ and we can consider a hot and a cold average velocity,
$v_h$ and $v_c$ respectively.  
In the limit of low $T$ and $w\ra0$, the DC-conductivity is given by
\be \label{DCstep}
\lim_{w\ra0}\sigma^{xx}=e^2\left[v_c^2 \tau_c N_c(\e_F)+ v_h^2 \tau_h N_h(\e_F)\right].
\ee
The quantity $N_c(\e_F)$ is the density of states at the Fermi level in the cold region, 
$N_c(\e)=\sum_{\v{k}}\d(\e_{\v{k}}-\e)\Phi_{\v{k}}$,
while $N_h(\e_F)$ is the same quantity in the hot region, with $N_h(\e)=\sum_{\v{k}}\d(\e_{\v{k}}-\e)\Psi_{\v{k}}$. Fig. 
\ref{figdos} shows the density of states in the hot (upper panel) and in the cold region (lower panel) as a function of the
size of the hot region (determined by the angle $\th$) in the limit $w\ra0$, obtained using the band
dispersion given in Eq. (\ref{Norman}).
Eq. (\ref{DCstep}) shows that 
cold and hot regions are wired in parallel, while each region (cold/hot) is wired in series with 
the transition region (in the case $w=0$), as shown by the structure of $\tau_c$ and $\tau_h$ given in Eq. (\ref{relaxtimes}).
In the limits mentioned above the temperature dependence of the resistivity
up to second order in $T$ is given by: 
\begin{eqnarray} \label{DCrest}
\lim_{w\ra 0}\rho^{xx}&=&\frac{1}{e^2}\frac{(1-\alpha)\bar{c}}{v_c^2N_c(\e_F)} T\\ && \nonumber
\left[1 +      
\left(\frac{\alpha}{1-\alpha}\frac{\bar{a}}{\bar{c}}   
   -\frac{\bar{c}}{\bar{b}}\left(\frac{v_h}{v_c} \right)^2
\frac{N_h(\e_F)}{N_c(\e_F)} \right) T \right].
\end{eqnarray}
As both lifetimes $\tau_c$ and $\tau_h$ for $w=0$ diverge in the limit $T\ra 0$, the resistivity 
has no zero temperature offset, $\rho(T\ra 0)=0$. 
However an offset can be obtained by a finite width $w$ as discussed above.

\begin{figure} 
\centering \psfig{file=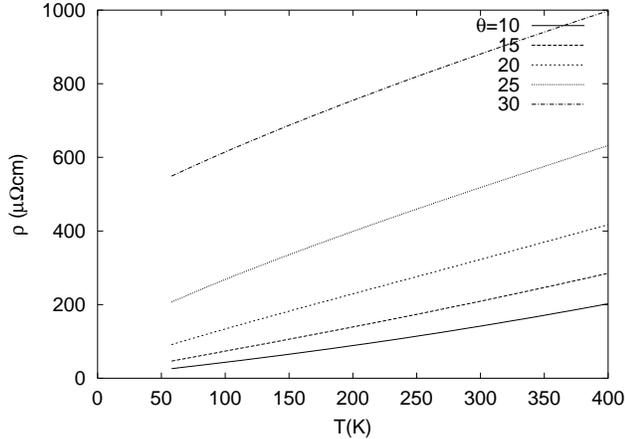, width=6cm,angle=-90}
\caption{Resistivity $\rho (T)$ for different sizes of the hot region (i.e. different $\th$) as a function of temperature.
The variation of the hot region changes the offset of the resistivity.  It can be 
seen that increasing the area of the hot region increases the residual resistivity $\rho_0$ and change slightly the power law
behavior $\rho (T) \sim T^\gamma$.  
The following parameters are used: ${\bar{a}}=48$, ${\bar b}=2$, ${\bar c}=7$ and $w=0.3$. $\th$ changes between $10^{\circ}$ and 
$30^{\circ}$.}
\label{theorestt}
\end{figure}

In our two-patch model we can associate the doping variation of the electronic properties of
the cuprates mainly with the variation of the area of the hot region 
and hence with the angle $\th$. Indeed, ARPES experiments on $Bi2212$ show that the region of the BZ where the spectral 
function is broad and no quasiparticle peaks are detectable increases as the doping is reduced.
A small segment of (quasiparticle) FS is observed approaching the metal-insulator transition \cite{Norman98}.
Therefore, we can translate this behavior of the FS and of the spectral function saying that when 
the doping is decreased, the angle $\th$ increases
(i.e., the size of the hot region increases).
Fig. \ref{theorestt} shows the change in resistivity with 
increasing angle $\th$; we obtain that the residual resistivity increases with increasing angle $\th$.

\begin{figure} 
\centering \psfig{file=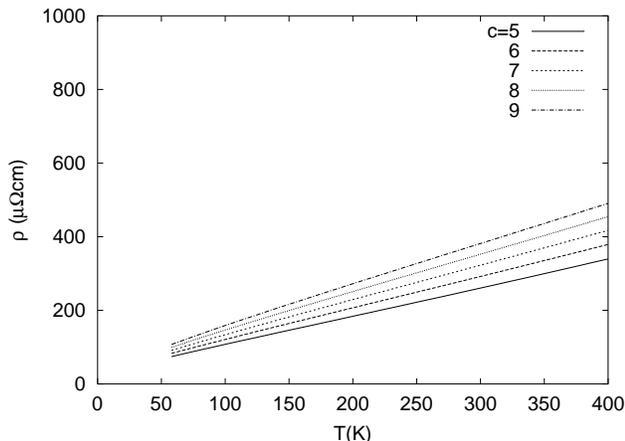, width=6cm,angle=-90}
\caption{The amplitude of the scattering between the cold and the hot region, described by $\bar{c}$, determines the
slope of the resistivity. Note that $\rho_0$ doesn't change. Increasing the inter-patch scattering 
determines an increasing in the slope of the DC-resistivity. 
The following parameters are used: $\bar{a}=48$, $\bar{b}=2$, $\th=20^{\circ}$ and $w=0.3$. $\bar{c}$ changes between $5$ and 
$9$.}
\label{theorestc}
\end{figure}

As can be seen in Eq. (\ref{DCrest}) the hot/cold scattering term $\bar{c}$ is the most important quantity in determining the 
slope of the resistivity respect to the other scattering amplitude $\bar{a}$ and $\bar{b}$.
The variation of the slope with $\bar{c}$ is reported in Fig. \ref{theorestc}. 
Note that 
a variation in $\bar{c}$ doesn't change the residual resistivity. 

Moreover, the slope of the resistivity is controlled by the single particle properties of the
cold region, in particular by the cold density of states $N_c(\e_F)$ and by the Fermi velocity $v_c^2$, and
by the area of the cold region $\alpha$, as obtained in Eq. (\ref{DCrest}). 

On the other hand, the influence of the prefactor 
$(1-\alpha)/(v_c^2N_c(\e_F))$ on the slope of the resistivity is small in comparison with the
influence of $\bar{c}$, as can be seen in Fig. \ref{theorestt} and Fig. \ref{theorestc}, because
of compensation effects: $v_cN_c(\e_F)$ is roughly $\theta$ independent and also the ratio $(1-\alpha)/v_c$
is expected to vary smoothly with $\theta$.

{\em Comparison with experiments.}
The linearity of the resistivity up to very high temperatures was found experimentally (see, e.g. \cite{Ando1999}).
Experiments show a quasi-linear temperature dependence of the resistivity 
$\rho^{xx}(T)\propto T^{\gamma}$, where in Ref. \cite{Ando1999}
it is shown that $\gamma$ increases with doping. For optimally doped systems $\gamma=1$, while for
overdoped systems a $\gamma > 1$ ($\gamma \simeq 1.5$) is observed, supporting a gradual recovering
of the Fermi liquid properties.
The doping dependence of the exponent $\gamma$ can be compared with our results
in Fig. \ref{theorestt}, showing that small values of $\th$ ($\th < 15^{\circ}$) give a resistivity
$\rho (T) \sim T^\gamma$ with $\gamma >1$. 
Moreover the residual resistivity $\rho_0=\rho(T=0)$ increases with underdoping 
and this is in agreement with the trend obtained in Fig. \ref{theorestt}.
Therefore, the two-patch model provides the possibility to change the doping mainly by changing the angle $\theta$,
besides considering the doping dependence of $\e_F$ and $v_F({\v{k}})$.
Note that the range of the change in $\rho$ with changing doping is observed in Ref. \cite{Ando1999} as well.
Fig. $1$ of Ref. \cite{Ando1999} shows that for underdoped $Bi2201$ (with a $La$ concentration of $x=0.66$) 
the offset $\rho _0 \simeq 150 \mu \Omega cm$, while for the overdoped compound (with $x=0.24$) $\rho_0\simeq 40 \mu \Omega cm$.
The increase in the slope and in the offset of the resistivity is also observed in $Y123$ as the 
doping is reduced \cite{Xiong}: the offset changes between $250 \mu \Omega cm$ and $20 \mu \Omega cm$ (Fig. 1(a) 
of Ref. \cite{Xiong}). The same range of variation is given in our results of Fig. \ref{theorestt} 
considering the range $10^{\circ}<\theta <25^{\circ}$.
Within the described change in $\theta$ a big variation of doping can be described.

In the following paragraph we report a
simple explicit comparison between the transport properties here evaluated and the 
experimental data via a fitting procedure.
The comparison here presented is limited to the $Bi$-based cuprates ($Bi2212$ and $Bi2201$) for 
which several ARPES and transport experimental data are available.

\begin{figure} 
\centering \psfig{file=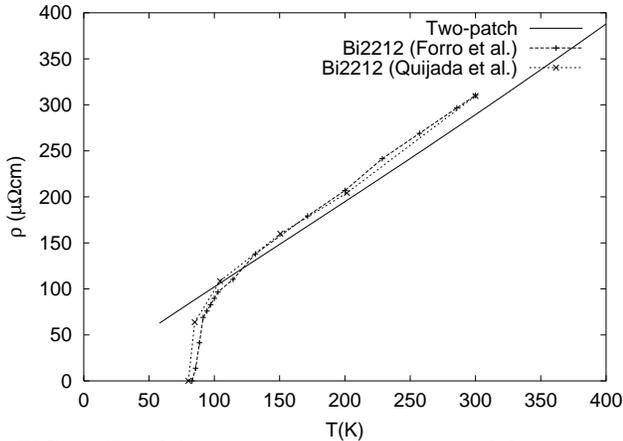, width=6cm,angle=-90}
\caption{Fit of the temperature dependence of the resistivity for the set of parameters given in the text. 
The experimental data are from an optimally doped $Bi2212$ samples from Ref. [1]
and from Ref. [43].}
\label{goodrest}
\end{figure}

The strategy of {\em fitting the experimental data} is the following.
We introduce a temperature scale $T_0$ upon which the term linear in $T$ in the scattering rate in the 
cold region dominates. This guarantees us that the resistivity is linear up to this temperature 
$T_0$. We choose a temperature $T_0$ that has a value of $\approx 1000K$. The 
relations $\bar{a}=(1-\alpha)\bar{c}/(\alpha T_0)$ and $\bar{b}=\alpha \bar{c}T_0/(1-\alpha)$ allows us to obtain first values
of $\bar{a}$ and $\bar{b}$ for given $T_0$ and $\bar{c}$. The width of the transition region $w$ is fixed mainly by the 
magnetoresistance as shown in section III-D. 
Starting with different angles $\th$ ({\em i.e.} different sizes of the hot region), we try to get a 
good fit of Hall-angle data as shown in section III-C. 
Then we try to fit the Hall-angle and the slope of 
the resistivity for different combination of $\th$ and $\bar{c}$. 
It turns out that the angle $\th \approx 20^{\circ}$ and the value $\bar{c} \approx 6.5$ gives a
good fit of the Hall-angle and of the slope of the resistivity. 
After that, we change the scattering amplitude $\bar{b}$ in the hot region. 
Increasing $\bar{b}$ leads to an higher resistivity and thus the parameter $\bar{b}$ allows us to adjust the 
offset of the resistivity. In this manner we can fix the parameters $w, \th, \bar{c}, \bar{b}$. 
The last parameter that is to be fixed is $\bar{a}$. The freedom for the parameter $\bar{a}$ is not big, 
in order to maintain the linearity of the resistivity. Thus the initial condition we have chosen for $\bar{a}$ can remain valid. 
We obtain a reasonable fit for the different transport quantities with the following values of the parameters:
$w=0.20$, $\bar{a}=60$, $\bar{b}=2.1$, $\bar{c}=7.0$ and $\th=20^{\circ}$.  
In Fig.\ref{goodrest} we report the resistivity as a function of temperature evaluated by Eq. (\ref{sigmaxx})
with the set of parameters given above and we compare our results with the resistivity 
measured in $Bi2212$ at optimum doping as given in Ref. \cite{Forro1990} and in Ref. \cite{Quijada1994}. 

\subsection{Thermoelectric power}

{\em Theoretical results.}
To evaluate the thermoelectricpower (TEP), which is the quantity measured in experiments, 
we first compute the longitudinal thermopower $S^{xx}$ given in Eq. (\ref{thermopower}).
The thermopower has almost the same expression as the DC-conductivity 
besides the extra factor of $\e_{\v{k}}$ and an extra minus in the sum, and it is given by  

\be \label{tepxx}
S^{xx}= -2e\sum_{\v{k}}\left(-\frac{\del f_{\v{k}}^0}{\del \e_{\v{k}}}\right)
\e({\v{k}})(v_{\v{k}}^x) ^2 \tau_{\v{k}}.
\ee
A Sommerfeld-expansion of Eq. (\ref{tepxx}), in the limit of $w=0$ and at low temperature, gives 
\be
\label{SommerTP}
\lim_{w\ra0}S^{xx}= -e\frac{\pi^2}{3}(k_BT)^2\left[v_c^2\tau_c N'_c(\e_F) + v_h^2\tau_h N'_h(\e_F)\right]  
\ee
with the first derivatives of the density of states $N'_{c,h}(\e_F)$ 
evaluated at the Fermi level in the cold and in the hot regions. We found that $S^{xx}$ has a smooth dependence 
on $w$ and hence the limit mentioned above is valid also for small values of the width $w$.
As in the case of conductivity, because $\tau_c>>\tau_h$ and $v_c>>v_h$,
the main contribution to the thermopower comes from the cold region. 
It is interesting that the Fermi level is below the peak in the cold density of states for $\th > 10^{\circ}$ 
(see Fig. \ref{figdos} (lower panel)). 
The peak in $N_c(\e)$ is exactly at the Fermi level for $\th \approx 10^{\circ}$. 
The TEP, as measured in the experiments mentioned above, is defined as  
\be\label{TEP}
TEP=\frac{S^{xx}}{T\sigma^{xx}},
\ee
where $\s^{xx}$ is given in Eq. (\ref{sigmaxx}). Again the cold region has the main
influence. The expression of the TEP in the limit of $w=0$ and at low temperatures is given by

\begin{eqnarray}\label{limitTEP}
\nonumber TEP&=&-\frac{\pi^2}{3 e}k_B^2 T\frac{ N'_c(\e_F)+[(v_h^2\tau_h)/(v_c^2\tau_c)]N'_h(\e_F) }{  
 N_c(\e_F)+[(v_h^2\tau_h)/(v_c^2\tau_c)] N_h(\e_F)} \\ &&
 \approx -\frac{\pi^2}{3 e}k_B^2 T\frac{ N'_c(\e_F)}{N_c(\e_F)}, 
\end{eqnarray}
where the last relation is obtained in the limit $\tau_c>>\tau_h$ and $v_c>>v_h$.
Therefore the ratio between the derivative of the density of states and the density of states evaluated at the Fermi level
in the cold region is the main quantity that determines the slope of the
TEP at low temperature. Increasing the temperature above $200 K$, we have verified that the contribution of the hot
region to the TEP becomes comparable to the contribution from the cold one.
\begin{figure} 
\centering \psfig{file=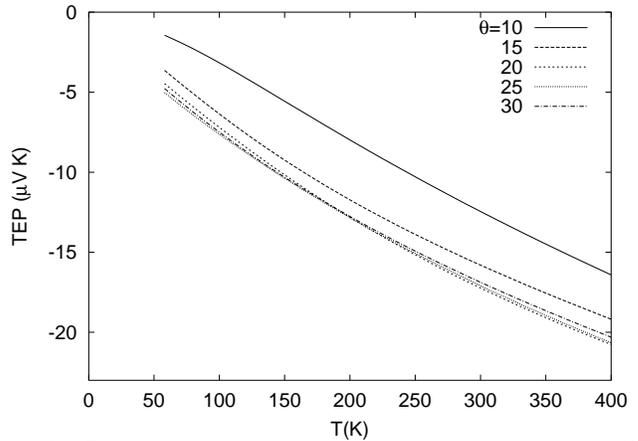, width=6cm,angle=-90}
\caption{Temperature dependence of the thermoelectric power for different 
angles $\th$. The following parameters are
used: $\bar{a}=48$, $\bar{b}=2$, $\bar{c}=7$ and $w=0.3$. $\th$ changes between $10^{\circ}$ and 
$30^{\circ}$.}
\label{theotept}
\end{figure}
Note that in the case of the TEP the offset doesn't change varying $w$. We found that 
the TEP is the most sensitive quantity to a variation of the electronic structure,
in agreement with Ref. \cite{McIntosh1996}. This is done by changing the 
hopping parameters $c_1$ and $c_2$ in Eq. (\ref{Norman}). 
Fig. \ref{theotept} shows the TEP obtained from the two-patch model for different angles $\th$.
For $\th\ge10^{\circ}$ the TEP is negative and  
the slope of the TEP at low temperature changes slightly with different angles $\th > 10^{\circ}$. 
We found that, as predicted in the limit given in Eq. (\ref{limitTEP}), the TEP 
is almost not affected by other parameters than $\th$. Nonetheless we got the 
largest effect on it from the scattering in the hot region. Again the slope doesn't change so much and the offset 
changes also slightly when the hot-scattering $\bar{b}$ is changed of a factor 5, as shown in Fig. \ref{theotepb}. 

\begin{figure} 
\centering \psfig{file=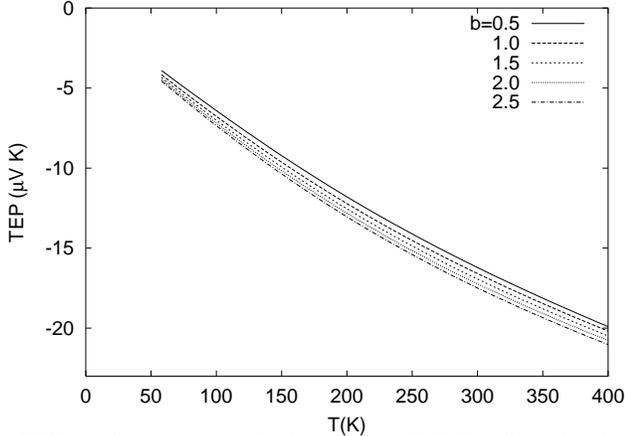, width=6cm,angle=-90}
\caption{Scattering in the hot region slightly affect the slope of the TEP and it has a weak influence 
on its offset. An increase in the hot-region scattering lowers the TEP and makes it even more negative.
The following parameters are used: $\bar{a}=48$, $\bar{c}=7$, $\th=20^{\circ}$ and $w=0.3$. $\bar{b}$ changes between $0.5$ and 
$2.5$.}
\label{theotepb}
\end{figure}

\begin{figure} 
\centering \psfig{file=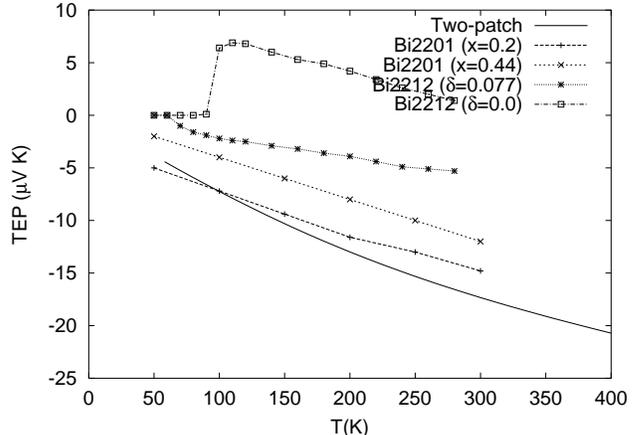, width=6cm,angle=-90}
\caption{Comparison between the TEP evaluated by the two-patch model and 
the thermoelectric data given by McIntosh {\em et al.} in Ref. [3]
(material: $Bi2201$ at optimum doping (x=0.44) and overdoping (x=0.2)) 
and by Obertelli {\em et al.} in Ref. [46]. 
(material: $Bi2212$ at optimum doping ($\delta=0.0$) and overdoping ($\delta=0.077$)).}
\label{goodtep}
\end{figure}

{\em Comparison with experiments.}
In Fig.\ref{goodtep} we report the TEP evaluated by Eq. (\ref{tepxx})
with the {\em same set} of parameters used for the resistivity
($w=0.20$, $\bar{a}=60$, $\bar{b}=2.1$, $\bar{c}=7.0$, $\th=20^{\circ}$)
and we compare our results with the TEP measured in $Bi2201$ at optimum and over doping 
as given in Ref. \cite{McIntosh1996} and with the TEP measured in $Bi2212$ at optimum and overdoping
as given in Ref.\cite{Obertelli}.
The TEP obtained with our two-patch model is only in qualitative agreement with the TEP observed 
by experiments. In particular only the overdoped $Bi2201$ has a TEP with a temperature 
dependence close to the one evaluated within the two-patch model. Note also that the TEP measured
in overdoped $Tl2201$ cuprates has almost the same temperature dependence and magnitude
of overdoped $Bi2201$ (and hence close to the TEP of the two-patch model), as shown
in Fig. 1 of Ref.\cite{Obertelli}.
We attribute the quantitative discrepancy between the TEP measured in optimally
and underdoped cuprates to the fact that the TEP reflects the 
properties of the excitations away from the FS and hence the frequency dependence of the scattering
amplitude becomes important, as shown by DMFT studies \cite{Palsson}.
We deserve the inclusion of the frequency dependence of the scattering amplitudes for future investigation.

\subsection{Hall angle}

{\em Theoretical results.}
The leading order contribution to the Hall-conductivity $\sigma^{xy}$ is given by the second term
in the expansion of the operator $\hat{A}^{-1}$ respect to the (weak) magnetic field, $\hat{A}^{-1}
=-\hat{K}^{-1}\hat{M}_B\hat{K}^{-1}$. Inserting this term into Eq. (\ref{electricconduc}), we get the 
Hall-conductivity $\sigma^{xy}$. The bending term in $\hat{A}^{-1}$ is
\be\label{lorenz}
({\v{v}}_{\v{k}}\times {\v{B}}) \nabla_{\v{k}} =B(v_{\v{k}}^y\del_{k_x}- v_{\v{k}}^x\del_{k_y}),
\ee
that arises from the operator $\hat{M}_B$ considering a magnetic field perpendicular
to the $CuO_2$ planes. In this case, the formula for the Hall-conductivity is the following
\be \label{Hall}
\sigma^{xy}=-2\frac{e^3 B}{\hbar c}\sum_{\v{k}}\tau_{\v{k}}v_{\v{k}}^{x}\left[v_{\v{k}}^y
\del_{k_x}- v_{\v{k}}^x
\del_{k_y} \right]\tau_{\v{k}} v_{\v{k}}^{y}\left(- \frac{\del f_{\v{k}}}{\del \e_{\v{k}}} \right).
\ee
The partial derivatives of the relaxation time $\tau_{\v{k}}$ enter in Eq. (\ref{Hall})
and are given by $\del \tau_{\v{k}}/\del k_x= -\tau_{\v{k}}^2(\del \Phi_{\v{k}})/(\del k_x)
\left[C_{\Phi}-C_{\Psi}\right]$ and $\del \tau_{\v{k}}/\del k_y=
-\tau_{\v{k}}^2(\del \Phi_{\v{k}})/(\del k_y)\left[C_{\Phi}-C_{\Psi}\right]$ respectively 
(note that $\del \Phi_{\v{k}}/\del k_x=-\del \Psi_{\v{k}}/\del k_x$). The final expression
for the Hall-conductivity is 
\begin{eqnarray}\label{hallconduc}
\sigma^{xy}&=& -\frac{2e^3 B}{\hbar c}\sum_{\v{k}}\left(- \frac{\del f_{\v{k}}}{\del \e_{\v{k}}} \right)
\tau_{\v{k}}^2 v_{\v{k}}^{x}\left[v_{\v{k}}^y
\frac{\del v_{\v{k}}^{y}}{\del k_x}- v_{\v{k}}^x
\frac{\del v_{\v{k}}^{y}}{\del k_y} \right] \\&+&  \nonumber 
 \frac{2e^3 B}{\hbar c}\sum_{\v{k}}\left(- \frac{\del f_{\v{k}}}{\del \e_{\v{k}}} \right)
\tau_{\v{k}}^3 v_{\v{k}}^{x}(v_{\v{k}}^{y})^2
\frac{\del \Phi_{\v{k}}}{\del k_x}\left[C_{\Phi} - C_{\Psi}\right]    
\\ &-& \nonumber  \frac{2e^3 B}{\hbar c}\sum_{\v{k}}\left(- \frac{\del f_{\v{k}}}{\del \e_{\v{k}}} \right) \tau_{\v{k}}^3 
(v_{\v{k}}^{x})^2v_{\v{k}}^{y}
\frac{\del \Phi_{\v{k}}}{\del k_y}\left[C_{\Phi} - C_{\Psi}\right]. 
\end{eqnarray}
This expression contains two different powers of the scattering time, 
one $\propto \tau_{\v{k}}^2$ and another $\propto \tau_{\v{k}}^3$.
In the limit $w\ra0$ we can get some further insight in the problem of the Hall-conductivity. In the low temperature
limit the sum over $\v{k}$ is 
restricted on the FS. The Hall-conductivity contains in this limit derivatives of step functions and hence   
$\d$-functions are generated.
As a consequence we get $8$ points on the FS that contribute to the second and third term in Eq. (\ref{hallconduc}).
The temperature dependence of the relaxation time $\tau_0$ evaluated in the transition region hot/cold 
on the FS can be written, using in this region
$\Phi_{\v{k}}=\Psi_{\v{k}}=1/2$ and using Eq. (\ref{defrelax}), 
as $\tau_0=1/(2C_{\Phi}+2C_{\Psi})=1/(\bar{c}_0+\bar{c}_1T+\bar{c}_2T^2)$. 
In this limit Eq. (\ref{hallconduc}) can be written in a more compact form as
\begin{eqnarray*}  
\lim_{w\ra0}\sigma^{xy}&=&\sigma_1^{xy}+\frac{2e^3 B}{\hbar c}4\tau_0^3\left[C_{\Phi}-C_{\Psi} \right]\\ &&
\left\{
v_y(A)v_y(B)\left[v_y(B)-v_y(A)\right] \right\}[q-1]
\end{eqnarray*}
where $\sigma_1^{xy}$ is the first term in Eq. (\ref{hallconduc}), $A$ and $B$ label the contributing points on the FS 
in the first quadrant of the BZ
and $q$ is either $\tan \th$ or $\cot \th$ depending on $\th$ respect to the value $\th \simeq 10^\circ$. 
Note that we get an offset to $\s^{xy}$ from the last terms even in
the limit $w\ra0$. Increasing $w$, also the first term in (\ref{hallconduc}) contributes to the offset due to the same 
reasons given for the resistivity.
The first term in (\ref{hallconduc}) contains the customer contribution proportional to $\tau_{\v{k}}^2$
while the second term contains higher powers in $\tau_{\v{k}}$ and can originate deviations from the
$\tau_{\v{k}}^2$ behavior depending from the temperature and the scattering amplitudes.

\begin{figure}
\centering \psfig{file=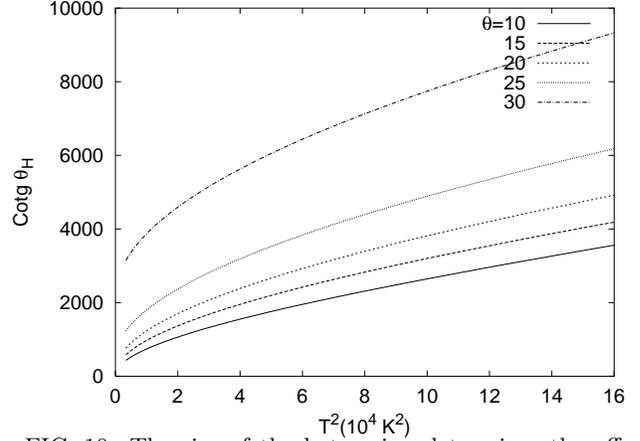, width=6cm,angle=-90}
\caption{The size of the hot region determines the offset of the Hall-angle. In the case of
$\th=30^{\circ}$, which is equivalent to $3/4$ of the first BZ
with hot character, the offset in the Hall angle is close to $50$. Note that we plot $\cot \th_H$ vs. 
$T^2$.  The following parameters are used $\bar{a}=48$, $\bar{b}=2$, $\bar{c}=7$ and $w=0.3$. 
$\th$ changes between $10^{\circ}$ and $30^{\circ}$. The magnetic field is $B=1 T$.}
\label{theohangt}
\end{figure}

\begin{figure} 
\centering \psfig{file=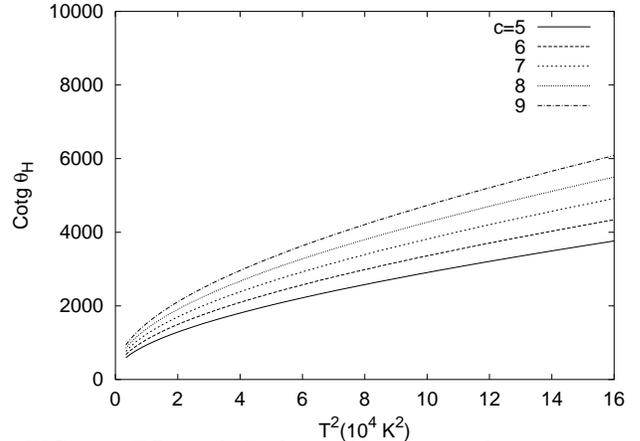, width=6cm,angle=-90}
\caption{Effect of the hot/cold inter-patch scattering on the 
Hall-angle. It turns out that $\bar{c}$ has an effect on the temperature dependence
of the Hall-angle, but no significant effect on the offset of the Hall-angle contrary to $\th$.
 The following parameters are used: $\bar{a}=48$, $\bar{b}=2$, $\th=20^{\circ}$ and $w=0.3$. $\bar{c}$ 
changes between 5 and 9. The magnetic field is $B=1 T$.}
\label{theohangc}
\end{figure}

The cotangent of the Hall-angle is defined as the ratio between the direct conductivity and the Hall conductivity,
\be\label{Hang}
\cot\th_H(T)=\frac{\sigma^{xx}}{\sigma^{xy}}.
\ee
A plot of $\cot\th_H$ vs. $T^2$ (see Fig.\ref{theohangt})
shows that the cotangent of the Hall-angle has a temperature 
dependence $\propto T^2$ in the high temperature regime ($T>200 K$), where we observe almost a straight line. 
It can also be seen in Fig. \ref{theohangt} that the range of temperature where $\cot\th_H \sim T^2$
increases with decreasing $\th$ (i.e. increasing doping). 
Konstantinovic {\em et al.} \cite{Konstantinovic2000} pointed out the strong influence of the anisotropy 
of the FS on the Hall-angle. We reproduced this
observation by changing the hopping parameters $c_1$ and $c_2$ in Eq. (\ref{Norman}) but we found that the effect of this 
change has even bigger effects on the TEP, as already discussed. 
Among the other parameters we found the strongest influence on $\cot\th_H(T)$ from the inter-patch scattering amplitude
$\bar{c}$. The effect of $\bar{c}$ on the Hall-angle is shown in Fig. \ref{theohangc}. 
It can be seen that 
this parameter changes the temperature range where $\cot\th_H \sim T^2$ only slightly   
and the offset is independent on $\bar{c}$.
The larger the value of $\bar{c}$ is the smaller the deviation of the Hall-angle from a straight line is, 
which means that increasing the cold/hot scattering has opposite effect as decreasing the angle $\th$.
On the other hand changing $\th$ (i.e. doping) effects the offset of the Hall-angle.
In the framework of the two-patch model we obtain a decreasing offset of the cotangent of the Hall-angle with increasing doping
and an increasing value of the power law coefficient $\gamma$ ($\cot\th_H \propto T^{\gamma}$) with increasing doping. 

{\em Comparison with experiments.}
As already discussed, the range of temperature where $\cot\th_H \sim T^2$
increases with decreasing $\th$ (i.e. increasing doping) (see Fig. \ref{theohangt}), which 
is contrary to the results given in \cite{Konstantinovic2000} and \cite{Ando1999}, but in agreement
with the results of Ref. \cite{Xiong}. Indeed a careful examination of various set of experimental
data for $\cot \th_H$ vs. $T^2$ indicates that the exact $T^2$ dependence does not extend over the
whole temperature range, and deviations are observed for $T<300 K$. As shown in Ref. \cite{Xiong}, e.g., in $Y123$ as $T_c$ is
lowered by underdoping, the range of the $T^2$ dependence of $\cot \th_H$ moves toward
higher temperatures (for $T_c=90 K$, the deviation start at $T=100 K$, while for $T_c=40 K$,
the deviation start at $T=170 K$). On the other hand
changing $\th$ (i.e. doping) effects the offset of the Hall-angle which was found in \cite{Ando1999}. 
Indeed in Fig. 3(b) of Ref. \cite{Ando1999} the change in the offset of the cotangent of the Hall
angle between underdoped (x=0.66) and overdoped (x=0.24) $Bi2201$ compounds measured at $T\simeq 280 K$
($T^2\simeq 8\cdot 10^4 K^2$) is of the order of 1300, a value which is compatible with our finding of
Fig. \ref{theohangt} ($\simeq 2000$) considering again $10^{\circ}<\th <25^{\circ}$ as in the case of
the resistivity discussed in Section III-A. In the case of $Y123$ the slope of $\cot \th_H$ vs. $T^2$
increases when the doping is reduced (almost $30\%$ of variation, when $T_c$ is reduced from $90 K$
for optimally doped to $40 K$ for an underdoped $Y123$, while in this case the offset increases only
slightly (see Fig. 1(e), Fig. 3 and Fig. 4 of Ref. \cite{Xiong}). Note that for $Y123$ a non monotonic
behavior of the slope of $\cot\th_H$ vs. $T^2$ as a function of doping is observed in a different set 
of measurements reported in Ref. \cite{Wuyts}.
This can be understood remembering that we increase the hot region increasing $\th$ (i.e. decreasing doping).
In the framework of the two-patch model we obtain a decreasing offset of the cotangent of the Hall-angle with increasing doping
and an increasing value of the power law coefficient $\gamma$ ($\cot\th_H \propto T^{\gamma}$) with increasing doping. 

\begin{figure} 
\centering \psfig{file=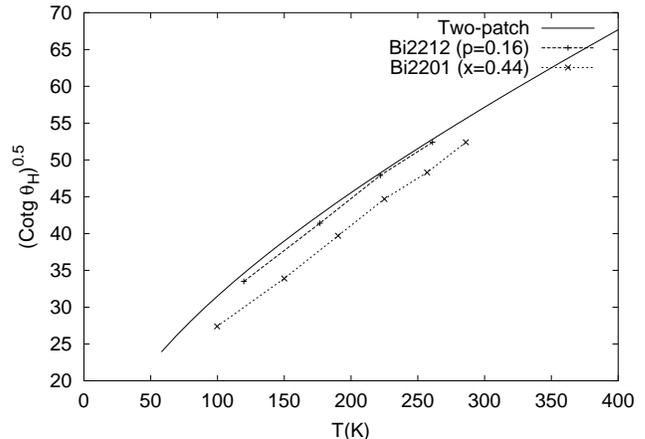, width=6cm,angle=-90}
\caption{Comparison between the experimental data for the Hall angle
and the Hall angle obtained from our two-patch model using the set of parameters given in the
text. The experimental data that are shown in the plot correspond to $Bi2212$ at optimum doping
from Ref. [6] and to $Bi2201$ at optimum doping from Ref. [7].}
\label{goodhang}
\end{figure}

In Fig.\ref{goodhang} we report the cotangent of the Hall angle evaluated by Eq. (\ref{Hang})
with the {\em same set} of parameters used for the resistivity and the TEP
($w=0.20$, $\bar{a}=60$, $\bar{b}=2.1$, $\bar{c}=7.0$, $\th=20^{\circ}$)
and we compare our results with the cotangent of the Hall angle measured in $Bi2212$ at optimum doping
as given in Ref. \cite{Konstantinovic2000} and in $Bi2201$ again at optimum doping as given in Ref. \cite{Ando1999}.
The agreement with the data obtained in $Bi2212$ is good and also a qualitative agreement is obtained if we
compare the results obtained with the two-patch model and the band structure of $Bi2212$ with the 
experimental data for $Bi2201$ at optimum doping.

\subsection{Magnetoresistance}

\begin{figure} 
\centering \psfig{file=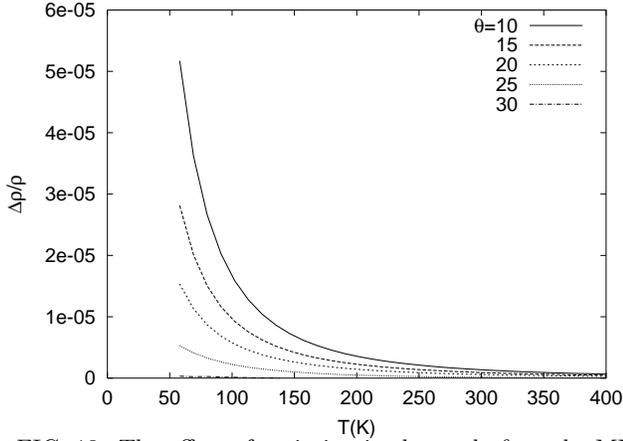, width=6cm,angle=-90}
\caption{The effect of variation in the angle $\th$ on the MR.
The MR decreases increasing the area of the hot region. 
In the limit where the whole BZ is hot ($\th=45^{\circ}$) the MR is zero. 
The parameters are $\bar{a}=48$, $\bar{b}=2$, $\bar{c}=7$ and $w=0.3$. $\th$ changes between $10^{\circ}$ and 
$30^{\circ}$. The magnetic field is $B=1 T$.}
\label{theomagnt}
\end{figure}

{\em Theoretical results.}
The  magnetoresistance (MR) is defined as the ratio of the variation 
in the resistivity in presence of a magnetic field to the resistivity  without magnetic field.
In the transverse geometry, with the magnetic field applied perpendicular to the $CuO_2$ planes and
the current measured parallel to the planes, the MR is given by
\be \label{MR}
 MR=\frac{\Delta \rho^{xx}(B)}{\rho^{xx}(0)}\approx-\frac{\Delta\sigma^{xx}(B)}{\sigma^{xx}(0)}-\tan^2\th_H .
 \ee
 In this geometry, the first extra contribution to the
 DC-conductivity, $\Delta \sigma^{xx}(B)$ is achieved by the third term in Eq. (\ref{Ainverse}), thus we have to insert
 $\hat{A}^{-1}=\hat{K}^{-1}\hat{M}_B\hat{K}^{-1}\hat{M}_B\hat{K}^{-1}$ into Eq. (\ref{electricconduc}).
 We get an expression for the correction to the conductivity given by
\begin{eqnarray} \label{delsigmaxx}
 \Delta\sigma^{xx}(B)&=& \nonumber
2e^2\sum_{\v{k},\v{k}^1,\v{k}^2,\v{k}^3,\v{k}^4,\v{k}^5}v_{\v{k}}^x\hat{K}^{-1}_{\v{k},\v{k}^1}
\hat{M}_{\v{k}^1,\v{k}^2}\hat{K}^{-1}_{\v{k}^2,\v{k}^3} \\ &&
\hat{M}_{\v{k}^3,\v{k}^4}\hat{K}^{-1}_{\v{k}^4,\v{k}^5}v_{\v{k}^5}^x \left(
-\frac{\del f}{\del \e_{\v{k}^5}}\right).
\end{eqnarray}
In this case it is necessary to consider partial derivatives of $\tau_{\v{k}}^2$ with respect to $k_x$ or $k_y$, which are
given by $\del \tau^2_{\v{k}}/\del k_x=-2\tau_{\v{k}}^3\del \Phi_{\v{k}}/\del k_x\left[C_{\Phi}-
C_{\Psi}\right]$ and $\del \tau^3_{\v{k}}/\del k_x=-3\tau_{\v{k}}^4(\del \Phi_{\v{k}})/(\del k_x)\left[C_{\Phi}-
C_{\Psi}\right]$. Finally we obtain $\Delta\sigma^{xx}(B)$ as:
\begin{eqnarray} \label{fullMR}
\nonumber \Delta &\sigma^{xx}&(B)=2\frac{e^4 B^2}{(\h c)^2}\sum_{\v{k}'}\left(
-\frac{\del f}{\del \e_{\v{k}'}}\right)\tau_{\v{k}'} v_{\v{k}'}^x
\left[v_{\v{k}'}^y\del_{k'_x}-v_{\v{k}'}^x\del_{k'_y}\right] \\
&&     \left\{\tau_{\v{k}'}^2 \left[v_{\v{k}'}^y\del_{k'_x}v_{\v{k}'}^x-v_{\v{k}'}^x\del_{k'_y}
v_{\v{k}'}^x \right] \right. \\&& \nonumber\:+ \left.
\tau^3_{\v{k}'} \left(
C_{\Phi}-C_{\Psi}\right)\left[
 (v_{\v{k}'}^x)^2  \frac{\del \Phi_{\v{k}'}}{\del k'_y}
-v_{\v{k}'}^xv_{\v{k}'}^y \frac{\del \Phi_{\v{k}'}}{\del k'_x}\right] \right\}.
\end{eqnarray}
We get three contributions to $\Delta \sigma^{xx}(B)$ with
different powers of $\tau_{\v{k}}$, $\propto \tau_{\v{k}}^3$, 
$\propto \tau_{\v{k}}^4$ and $\propto \tau_{\v{k}}^5$.
This formula is inserted in Eq. (\ref{MR}) together with the Hall-angle computed in the previous 
subsection.  We are now able to compute the 
MR for different parameters.
As shown in Fig. \ref{theomagnt}, a change in the angle $\th$ (and 
hence in doping) has a sizeable effect on the MR and decreasing the angle $\th$ ({\em i.e.} increasing doping) 
increases the MR in our model.

\begin{figure} 
\centering \psfig{file=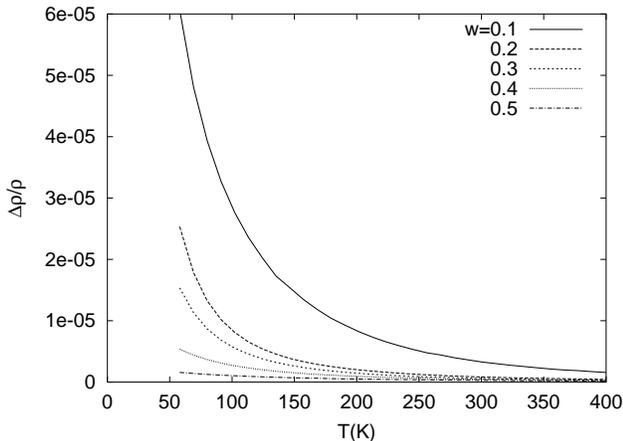, width=6cm,angle=-90}
\caption{The MR diverges in the limit $w\rightarrow0$. The influence 
of $w$ on the MR is very strong, so the MR allows us to fix the parameter $w$ quite well as other quantities
don't depend strongly on $w$. We used the parameters $\bar{a}=48$, $\bar{b}=2$, $\bar{c}=7$ and $\th=20^{\circ}$ in the plot. 
$w$ varies between $0.1$ and 
$0.5$. The magnetic field is $B=1T$.}
\label{theomagnw}
\end{figure}

The effect of the transition width $w$ is studied in Fig. \ref{theomagnw}. 
It can be seen that this 
quantity becomes more and more important the smaller it becomes. A big difference in the MR can be 
observed between $w=0.1$ and $w=0.2$, which is in agreement with the results for the MR 
evaluated within a cold spots model by Zheleznyak {\em et al.} reported in Ref. \cite{Zheleznyak1999}.

\begin{figure} 
\centering \psfig{file=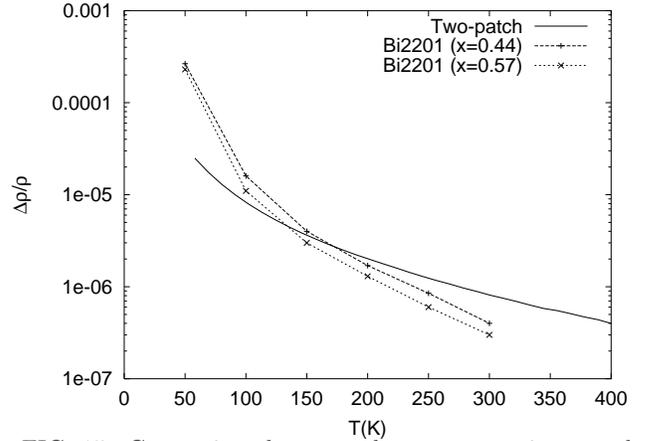, width=6cm,angle=-90}
\caption{Comparison between the magnetoresistance obtained from the two-patch model and 
the orbital magnetoresistance of an optimally doped $(x=0.44)$ 
and slightly underdoped  $(x=0.57)$ $Bi2201$ from Ando {\em et al.} [7].}
\label{goodmagn}
\end{figure}

{\em Comparison with experiments.}
Ando {\em et al.} reported that the orbital MR increases 
with increasing doping, as shown in Fig. 5 of Ref. \cite{Ando1999}. (The orbital contribution
to the MR can be obtained by the transverse component of MR subtracting the longitudinal component,
eliminating in this way the contribution of the spins to the MR.)
This experimental result is derived in our model.
In Fig. \ref{goodmagn} we report the magnetoresistance evaluated by Eq. (\ref{MR})
with the {\em same set} of parameters used for the other transport properties above discussed 
($w=0.20$, $\bar{a}=60$, $\bar{b}=2.1$, $\bar{c}=7.0$, $\th=20^{\circ}$) together with $B=1 T$
and we compare our results with the orbital magnetoresistance measured in $Bi2201$ at optimum 
doping and slightly underdoping
as given in Ref. \cite{Ando1999}. (MR data for $Bi2212$ are not yet available to our knowledge).
For the high doping level and range of temperature here considered the longitudinal MR is an order of
magnitude smaller than the transverse MR and its contribution to the orbital MR is therefore small.

While the order of magnitude and the qualitative temperature dependence of the MR evaluated with the two-patch 
model agree with the MR data for $Bi2201$, a quantitative discrepancy is observed taking fixed the set of parameters
we used to evaluate the previous transport properties. On the other hand a change in $w$ from 
$w=0.2$ to $w=0.1$ can increase the MR curve shown in Fig. \ref{goodmagn}, leading to a better agreement
with the low temperature MR data for $Bi2201$. 
Note that the width of the transition region $w$ is associated to the
properties of the electronic scattering and its value can be material dependent, even in presence
of a similar FS and band structure. 
A temperature dependence of the coefficient $w$ seems necessary to improve the fit of the MR data.
MR data for optimally doped $Bi2212$ could permit a more careful comparison with
the prevision of the two-patch model here presented.

\subsection{Thermal Hall conductivity}

{\em Theoretical results.}
The thermal-Hall conductivity is evaluated using Eq. (\ref{thermconduc}). 
In this case the operator $\hat{A}^{-1}$ has the form $\hat{K}^{-1}\hat{M}_B\hat{K}^{-1}$.
Replacing $e^2\rightarrow \e_{\v{k}}^2$ in Eq. (\ref{hallconduc}) gives us 
the result for $\kappa^{xy}$.
\begin{eqnarray} \label{numthermalhall}
\nonumber \k^{xy}&=& -\frac{2eB}{\hbar c}\sum_{\v{k}}\e_{\v{k}}^2\left(- \frac{\del f_{\v{k}}^0}{\del \e_{\v{k}}} \right)
\tau_{\v{k}}^2 v_{\v{k}}^{x}\left[v_{\v{k}}^y
\frac{\del v_{\v{k}}^{y}}{\del k_x}- v_{\v{k}}^x
\frac{\del v_{\v{k}}^{y}}{\del k_y} \right]\\&& +  \nonumber 
\frac{2e B}{\hbar c}\sum_{\v{k}}\e_{\v{k}}^2\left(- \frac{\del f_{\v{k}}^0}{\del \e_{\v{k}}} \right)
\tau_{\v{k}}^3 v_{\v{k}}^{x}(v_{\v{k}}^{y})^2
\frac{\del \Phi_{\v{k}}}{\del k_x}\left[C_{\Phi} - C_{\Psi} \right]    
\\ && -    \frac{2e B}{\hbar c}\sum_{\v{k}}\e_{\v{k}}^2\left(- \frac{\del f_{\v{k}}^0}{\del \e_{\v{k}}} \right) \tau_{\v{k}}^3 
(v_{\v{k}}^{x})^2v_{\v{k}}^{y}
\frac{\del \Phi_{\v{k}}}{\del k_y}\left[C_{\Phi} - C_{\Psi} \right]. 
\end{eqnarray} 

\begin{figure} 
\centering \psfig{file=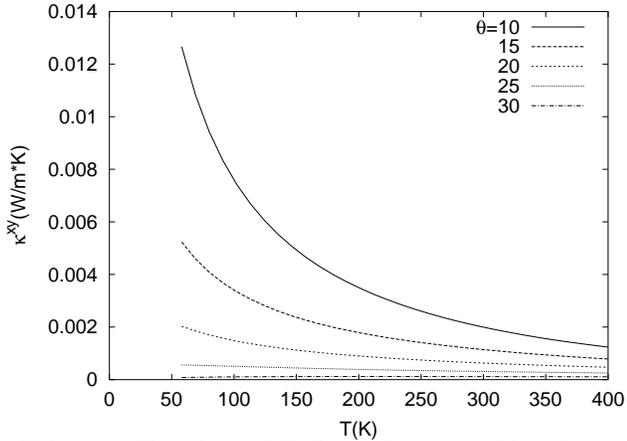, width=6cm,angle=-90}
\caption{The thermal-Hall conductivity (like electrical transport properties) 
is mostly influenced  by the cold region. 
We use the following parameters: $\bar{a}=48$, $\bar{b}=2$, $\bar{c}=7$ and $w=0.3$. $\th$ changes between $10^{\circ}$ and 
$30^{\circ}$.}
\label{theothallt}
\end{figure}

In Fig. \ref{theothallt} the thermal-Hall conductivity is reported as a function of the temperature. 

\begin{figure} 
\centering \psfig{file=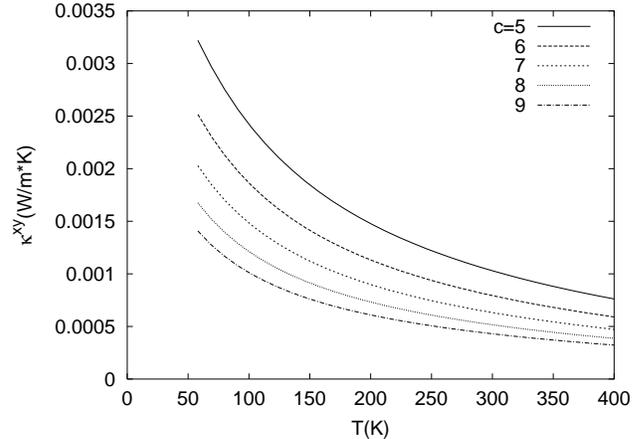, width=6cm,angle=-90}
\caption{A change in $\bar{c}$ shifts the thermal Hall conductivity: an increasing of the inter-patch scattering 
leads to a decreasing of the thermal-Hall conductivity. The following parameters are used: 
$\bar{a}=48$, $\bar{b}=2$, $w=0.3$ and $\th=20^{\circ}$. $\bar{c}$ changes between $5$ and 
$9$.}
\label{theothallc}
\end{figure}

The various curves correspond to different values of the angle $\th$ and we obtain a large increase of
$\k^{xy}$ as the angle $\th$ is decreased. As in the case of the other transport properties, the cold
patches has the main influence in determining $\k^{xy}$, even if the contribution to $\k^{xy}$ from the
hot patches is sizeable as in the case of the TEP. The role of the inter-patches coupling in $\k^{xy}$ is studied
changing the scattering amplitude $\bar c$ and the results are reported in Fig. \ref{theothallc}, showing that
increasing $\bar c$ tends to suppress $\k^{xy}$ in a sizeable way. The same behavior is observed considering the
role of the hot patches, changing the scattering amplitude $\bar b$.
 
Considering the qualitative correspondence between decreasing doping and increasing $\th$ discussed above,
the two-patch model suggests that in underdoped cuprates the thermal-Hall conductivity
should be strongly suppressed with respect to the one measured in optimally and overdoped cuprates.
This conclusion is also supported  by the fact that the electronic scattering in the hot region
(and hence $\bar b$) should increase in the underdoped regime due to the proximity to the
antiferromagnetic phase.

\begin{figure} 
\centering \psfig{file=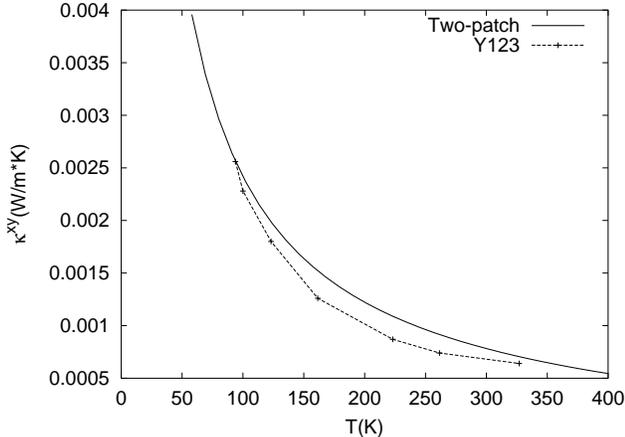, width=6cm,angle=-90}
\caption{Thermal-Hall conductivity $\kappa^{xy}$ vs. temperature obtained from the two-patch model
with the same set of parameters given in the text.
The temperature dependence of $\kappa^{xy}$ agrees quite well as regard the magnitude with the data for 
$\kappa^{xy}$ measured for an optimally doped YBCO [8].}
\label{goodthall}
\end{figure}

{\em Comparison with experiments.}
In Fig. \ref{goodthall} we report the thermal-Hall conductivity evaluated by Eq. (\ref{numthermalhall})
with the {\em same set} of parameters used for the other transport properties above discussed 
($w=0.20$, $\bar{a}=60$, $\bar{b}=2.1$, $\bar{c}=7.0$, $\th=20^{\circ}$).
Again, experimental data for $\kappa^{xy}$ for the $Bi$-based cuprates are to our knowledge not yet available.
The temperature dependence of $\kappa^{xy}$ agrees quite well as regard the magnitude with the data for 
$\kappa^{xy}$ measured for an optimally doped YBCO \cite{Zhang2000}, even if this material has some differences in
the band structure and FS respect to the $Bi$-based cuprates. 

\section{Conclusion and Discussion}
\label{s5}

Normal state transport properties of cuprate superconductors have been studied 
using a semi-classical approach based on the linearized Boltzmann equation (BE).
The probability of scattering between the electrons of the conduction band and an effective 
collective mode is assigned by a scattering matrix defined on a 
Brillouin zone (BZ) divided in two kind of patches. One kind of patches, centered in the $M$ points of the
BZ, contains {\em hot states}, which are strongly interacting, and is characterized by a low
Fermi velocity and an high local density of states. The other kind of patches, centered in the nodal
points along the BZ diagonals, contains {\em cold states}, which are weakly interacting, and is 
characterized by an high Fermi velocity and a low local density of states.
In the multi-patch model the scattering matrix is assumed to be a sum of separable terms (in $\v{k}$ and $\v{k'}$)
having coefficients with a temperature dependence which can be non conventional, according to the
properties of the scattering.
In the case of the {\em two-patch model}, the scattering matrix is a sum of a term describing the scattering 
between the electrons inside the hot patches with a large amplitude and
temperature independent, a term describing the scattering inside the cold patches 
with a smaller amplitude and dependent on the temperature as $T^2$, and a 
term including in a symmetric way the inter-patch scattering, with a linear
temperature dependence. With this phenomenological choice of the scattering matrix, the BE is
exactly solvable, and all the transport properties (at least in the weak field regime)
can be evaluated. The resulting scattering amplitude $1/\tau_{\v{k}}$ is strongly momentum
dependent (in particular along the Fermi surface) 
and the low temperature behavior is always non Fermi liquid, with a linear temperature
dependence in the cold patches and a constant in the hot patches, as suggested by recent ARPES experiments
performed on $Bi2212$ \cite{Valla}. 
The deviation of the distribution function from the equilibrium is strongly suppressed in the hot patch 
because of the strong scattering, while in the cold patch it has a sizeable value, 
because of the weaker scattering, giving the largest contribution to transport. 

The two-patch model here introduced has similarities as well differences with the model 
of Hlubina and Rice \cite{Hlubina1995}. Hlubina and Rice consider a model
where the fermions are scattered by an antiferromagnetic spin fluctuation. The
propagator of the spin fluctuations is peaked at the antiferromagnetic wave-vector $\v{Q}\equiv (\pi;\pi)$ and hence 
couples mainly states around the $M$ points, giving rise
to hot spots, while the states around the nodal points are weakly coupled, giving rise to cold regions.
The scattering matrix correspondent to this interaction has two different temperature
regimes: ($i$) the low temperature regime, where the scattering between the hot states has a $\sqrt{T}$ behavior,
the scattering between the cold states has a $T^2$ Fermi liquid behavior and the scattering between hot/cold states
has again a $T^2$ behavior; ($ii$) the high temperature regime is instead similar to our two-patch model,
having a constant scattering between the hot states, a quadratic ($\sim T^2$) scattering between the cold states 
and a linear in $T$ scattering between hot/cold states. Therefore, the model of Hlubina and Rice gives
at low temperature a $T^2$ (Fermi liquid) temperature dependence of the resistivity, while our two-patch model
gives a linear in $T$ (non-Fermi liquid) behavior of the resistivity even at low temperature.

An additive two-lifetime model, with similarities to our two-patch model, as been previously
proposed \cite{Zheleznyak1999}. The results obtained within this model are based on a peculiar
Fermi surface, characterized by large flat regions parallel to the $\Gamma M$ directions
having a short relaxation time (hot regions) and small sharp corners around the nodal
points along the $\Gamma Y(X)$ directions having a long relaxation time (cold regions).
Moreover the band structure considered in Ref.\cite{Zheleznyak1999} is such that the Fermi velocity is large 
in the hot region and small in the cold region. The authors suggest that this peculiar
single particle properties are typical for $YBCO$ at optimum doping. ARPES experiments
for $Bi2212$ and $Bi2201$ do not support this picture, and in particular the 
ratio between the Fermi velocity is the opposite, having large Fermi velocity
in the cold region and small (almost undefined) Fermi velocity in the hot region.
The flat region of Fermi surface observed in $Bi$-based compounds is also much smaller
than the one proposed in Ref.\cite{Zheleznyak1999}. 
This flat shape of the Fermi surface seems more appropriate for $La$-based underdoped cuprates,
where stripe correlations can select preferred directions on the Fermi surface \cite{Ino}.
The regions of the Fermi surface controlling the
transport properties in the additive two-lifetime model and in our 
two-patch model are different. In particular the conductivity is controlled in the first
case by the flat (hot) regions, while in our case is controlled by the curved (cold) regions,
mainly because of the completely different ratio between the Fermi velocities.
In the two-lifetime model the scattering amplitude in the hot region has a linear temperature
dependence and a quadratic (Fermi liquid like) temperature dependence in the cold region.
Therefore in this model the resistivity is linear in temperature, while the cotangent of the Hall angle
is roughly quadratic in temperature, being the Hall conductivity mainly controlled by the
regions of Fermi surface with sizeable curvature (as the corners).
In our approach both the resistivity and the Hall conductivity are mainly controlled
by the cold regions of the Fermi surface, where, as already discussed, the
scattering amplitude has a non Fermi liquid character at low temperature.

A further comparison can be done with the model proposed by Ioffe and Millis \cite{Millis1998},
where the scattering amplitude is a sum of two terms, one temperature independent
with an angle dependence along the Fermi surface 
with a deep (quadratic) minimum along the diagonal directions ($\Gamma Y(X)$) and the other Fermi liquid-like, with a quadratic
temperature dependence. At low temperature this model gives a large scattering
amplitude constant in temperature in the (large) hot region and a scattering amplitude
with a $T^2$ Fermi liquid temperature dependence in the (small) cold region (cold spots). The linear resistivity
is obtained in this model because the conductivity is dominated by a small (cold) region which
has a length proportional to the temperature, where the scattering is Fermi liquid-like.
In our two-patch model the area (and hence the length) of the cold region is considered to be temperature
independent and a direct comparison with the model of Ioffe and Millis is not possible.
The form of the scattering amplitude proposed in this cold spots model is similar to the form
proposed by Valla {\em et al.} \cite{Valla} to describe the temperature and momentum dependence of the
width of the (quasiparticle) peak in the ARPES spectra. On the other hand the experiments are
consistent with a linear temperature dependence of the width around the zone diagonals. 
Interestingly, a quasiparticle peak is present in the ARPES spectra for all the Fermi wave-vector
along the curved area of the Fermi surface, which is clearly not only a sharp corner, but
a sizeable fraction (almost the half) of the whole Fermi surface.  
Of course a direct comparison between the scattering amplitude or the scattering matrix elements
proposed in the various models and the ARPES line-shape is not possible and only some
qualitative understanding can be obtained from ARPES experiments without a microscopic theory
which is able to connect two-particle and single-particle properties.

Another model for magnetotransport in cuprates has been recently proposed by Varma and Abrahams \cite{Varma}.
This model combines the marginal Fermi liquid hypothesis for the inelastic scattering rate (linear in T), with
the hypothesis that small-angle forward scattering is acting in the cuprates due to the scattering
of the electrons of the $CuO_2$ planes with the out of plane impurities; the forward scattering term in
the scattering rate results temperature independent and strongly anisotropic. The total form of the
scattering rate is again similar to the form proposed by Valla {\em et al.} \cite{Valla} to fit
the ARPES spectra of $Bi2212$ as already discussed. Solving the BE, the authors show that forward scattering
is responsible for a correction term in $\tan \th_H$ respect to the customary contribution and this term
has the temperature dependence of the resistivity squared. In our approach the correction term corresponds
to the second and third terms in Eq. (\ref{hallconduc}), proportional to $\tau^3$. On the other hand 
in our two-patch model we obtain a finite contribution from the first term, proportional to 
$\tau^2$ (the customary term), which is of the order or larger of the other terms. A detailed comparison between
the two approaches requires to consider a five-patch model as discussed in Section II.

Our {\em systematic analysis} of normal state transport properties of cuprate superconductors,
including resistivity, thermoelectric power, Hall conductivity, magnetoresistance and 
thermal-Hall conductivity, permits to understand which is the role of the patch geometry, of the Fermi surface
and of the scattering matrix elements in determining the magnitude and the temperature dependence of
the various electrical and thermal transport properties. In particular, the linear temperature dependence of the 
resistivity is associated to the inter-patch scattering, and its slope is determined by the amplitude
of the inter-patch scattering but also by the single particle properties in the cold patch, as
the cold density of states evaluated at the Fermi level and the cold Fermi velocity. The Hall conductivity is also governed
by the cold patch and the interplay between the various power law coming from the scattering time
and from its partial derivatives gives the possibility to obtain a cotangent of the Hall angle
with a $T^{\gamma}$ behavior (with $\gamma>1.5$) in a range of temperature where the resistivity is linear.
The different power law behavior of resistivity and cotangent of the Hall angle indicates that the
momentum dependence of the scattering time along the Fermi surface plays an important role and
can originate a different characteristic scattering time for longitudinal and transverse transport.
The magnetoresistance is mainly determined at low temperature by the inter-patch region, and in particular
its expression contains the second derivative of the scattering time. Therefore, the transition between the hot and
the cold patches has to be smooth, in order to avoid a spurious divergence. 
Thermal transport is also considered in our systematic analysis. Thermoelectric power (TEP), as thermal-Hall conductivity,
are mainly determined by the cold patch in the low temperature regime, and it is interesting to note that the 
slope of the TEP is only controlled by the cold density of states and its derivative, giving a strong 
connection between the experimental slope of the TEP and the patch geometry.
Increasing the temperature ($T>200 K$), the hot patch starts to contribute to the thermal properties.

Finally, we present a tentative application of our two-patch model to the electrical and thermal 
transport properties of optimally and overdoped $Bi$-based cuprates. 
We use the electronic band structure and Fermi surface obtained by ARPES
experiments and we obtain other information by the temperature dependence of the ARPES lineshape.
The amplitudes of the intra-patch and inter-patch terms in the scattering matrix are fixed using 
the measurements of the resistivity, while the patch geometry is fixed by the TEP. Once all
the parameters have been fixed, Hall conductivity, magnetoresistance and thermal-Hall conductivity
are evaluated without any other assumptions and  a reasonable agreement between our two-patch model and 
experimental values is found. In conclusion, the two-patch model for the scattering process 
emerges as a minimal division of the Brillouin zone to account for the strong anisotropy of
the effective electron-electron interaction present in the cuprates, which is able to describe
the several anomalous temperature dependences 
observed in the normal state transport properties of cuprate superconductors.
The description of the transport properties here presented can be improved increasing the 
number of patches, e.g. to N$=5$ to include forward scattering processes.

The functional form of the scattering operator can be realized in C-DMFT calculations. 
Work to see if a microscopic model such as a Hubbard model for some choice of parameters
produces a temperature dependence close to our optimal fit of the data is currently
under investigation.

\acknowledgments
We appreciate valuable discussions with C. Castellani, M. Cieplak, M. Civelli, D. Drew and R. Raimondi. 
A. Perali acknowledges partial support from Fondazione "Angelo della Riccia". M. Sindel is grateful to DAAD for partial 
support.


\begin{references}
\bibitem{notatwogap} The hot/cold division of the BZ has also been used in the two-gap model to describe the pseudogap
opening in underdoped cuprates induced by pair fluctuations as discussed in Ref.\cite{Perali}.
\bibitem{notamp} The zero average of $C_{k,k'}v_k'$ over the BZ 
is no longer valid when a generic many-patch model is considered.
For instance, when forward scattering is included in $C_{k,k'}$, the scattering between different cold regions 
is not equivalent any more with the scattering inside a given cold region, and the average given above
is in general different from zero. The general solution for the multi-patch model given in 
Eqs. \ref{mpgk},\ref{tij},\ref{kkk} has to be considered.
\end{references}
\end{document}